# Concurrent Composition and Algebras of Events, Actions, and Processes

Mark BURGIN [a] and Marc L. SMITH [b]

[a] *Department of Computer Science, Univ. of California, Los Angeles*
*Los Angeles, California 90095, USA*
[b] *Department of Computer Science, Vassar College*
*Poughkeepsie, New York 12604, USA*

**Abstract.** There are many different models of concurrent processes. The goal of this work is to introduce a common formalized framework for current research in this area and to eliminate shortcomings of existing models of concurrency. Following up the previous research of the authors and other researchers on concurrency, here we build a high-level metamodel EAP (event-action-process) for concurrent processes. This metamodel comprises a variety of other models of concurrent processes. We shape mathematical models for, and study events, actions, and processes in relation to important practical problems, such as communication in networks, concurrent programming, and distributed computations. In the third section of the work, a three-level algebra of events, actions and processes is constructed and studied as a new stage of algebra for concurrent processes. Relations between EAP process algebra and other models of concurrency are considered in the fourth section of this work.

**Keywords:** event, action, process, signal, action composition, process composition, true concurrency, process algebra

## 1. Introduction: Problems of Concurrency and Related Research

The algebraic approach is a popular and effective way to study processes in multicomponent systems. To efficiently deal with such problems computer science developed a concurrency theory, which extensively and successfully utilizes algebraic structures and techniques. Researchers built many powerful models of concurrency: Petri nets, CSP, CCS, ESP, VCR, synchronization trees, labeled transition systems, and grid automata, to mention but a few.

The goal of this work is to introduce a common formalized framework for current research in this area and to eliminate shortcomings of existing models of concurrency. The main problem with the majority of these models is that they use an oversimplified model of time. Firstly, they consider only two types of time scales – linear time and branching time, while the system theory of time (Burgin, 2002) implies a necessity to have a more flexible representation of temporal concurrent processes. With the advent of multi-core processors, and the prevalence of cluster and grid computing, reasoning about true concurrency is no longer an avoidable concern. Secondly, all events and actions in these models of concurrency do not have duration, while in reality to build reliable technological systems, such as computers and networks, it is necessary to take into account duration of events and actions (Denning, 2007). Thirdly, it is generally assumed that moments of time when events occur are exactly specified and distinguishable from one another, while in reality researchers and engineers encounter an impossibility to distinguish two near-simultaneous events (Lamport, 1984; Denning, 2007). As a result, the arbitration problem is one of the most important for concurrent systems (Denning, 1985).

In addition, from computability theory, Rice's theorem (Rice, 1951) restricts us to observing input/output behavior of processes to reason about properties of computation. But in concurrent systems, observable input/output behavior extends to communication between processes, and process algebras such as CSP, CCS, and the Pi-calculus are built up from elemental notions of observable events. Algebraic laws prescribe a variety of compositional possibilities, processes are composed of events and other processes, and computational histories of processes are represented by traces of their observable events. Simplifying assumptions, such as instantaneous events (events without duration) and the interleaving of parallel events in computational traces, have proven successful to date, but we may now be confronting certain limitations of such assumptions. Reasoning about composition in general, but especially concurrent composition, remains one of our greatest challenges. One of the reasons is that, as experts claim, nontrivial concurrent programs based on threads, semaphores, and mutexes are incomprehensible to humans.

The extended view-centric reasoning (EVCR) model for concurrent processes, developed by Burgin and Smith (2006, 2006a), gives a flexible mathematical representation for temporal relations between such processes, but lacks a mathematical model of these processes, as well as their system representation. Here we enrich EVCR with such a model, constructing a high-level metamodel, EAP (event-action-process), for concurrent processes. As a base for this new component, we take the metamodel, ESP (event-signal-process), built by Lee and Sangiovanni-Vincentelli (1996). This metamodel gives an efficient hierarchy for mathematical representation of concurrent processes and encompasses a variety of models of concurrent systems. It is demonstrated in Lee and Sangiovanni-Vincentelli (1996) that ESP can represent Kahn process networks (Kahn, 1974), the CSP model of Hoare (1985), the CCS model of Milner (1989), dataflow process networks (Lee and Parks, 1995), discrete event simulators (Fishman, 1978), and Petri nets (Petri, 1962; Reisig, 1985).

EAP is inspired by CSP's observation-based reasoning capability. It is the next step in the evolution of CSP from a model that reduces concurrency to an interleaving of sequential events, to a model of true concurrency that provides abstractions that represent degrees of interleaving. EAP also is the next step in the evolution of ESP from a model in which events happen at a moment of time (point-like events) and time is absolute, to a model that provides abstractions to represent extended events and relativity of time. Event duration is important for efficiency of computation and communication and thus, is the base for computational complexity. As a result, a model of true concurrency should support abstractions for events whose occurrence partially overlap in time. Thus, the possibility of observed event simultaneity is not merely a Boolean proposition, but rather a continuum: events A and B may overlap entirely, partially, or not at all.

Denning (2007) recently described the choice uncertainty principle, a phenomenon that exists at every level of computing abstraction from the hardware up, as well as in human social situations. The choice uncertainty principle states "it is impossible to make an unambiguous choice between near-simultaneous events under a deadline." Denning goes on to ask, "How can computations choose reliably?" We believe EAP provides one path to begin reasoning about concurrency by reintroducing event duration into observation-based computational models.

EAP begins by deconstructing the notion of events, and continues by deconstructing observation, itself. In Section 2, several event axioms are given whose truth and meaning are evident in the real world, but these are not the axioms of events that appear in classic process algebras, such as CSP, CCS, and ESP. To achieve higher flexibility, in reflecting real concurrent processes and distributed systems, as well as concurrent programs, at first, we use semi-formal definitions, which allow later several exact formalizations. For instance, at first, we introduce a general concept of the temporal coordinate of an event, and only then we give three possible mathematical realizations: the temporal coordinate as a real number for point-like events, the temporal coordinate as a real interval for extended events, and the temporal coordinate as an ordered set (tuple) of real numbers for pointed events.

Events in the real world exist in space and time, but events are more typically abstracted to occur at a single point in time, without regard for location or duration. At its most fundamental level, an event represents change. EAP considers the meaning of an event from three perspectives: the change in the process that engaged in the event, the change in the environment in which the event occurred, and the change in the relationship between a process and its environment when an event occurs. In View-

Centric Reasoning (VCR), Smith (2000) considered multiple views of a computation; in EAP, we go so far as to consider different views of events, themselves.

EAP deconstructs observation by considering intermediate units of event composition. Process algebras typically define processes in terms of event composition, with primitives that permit sequential composition, choice, and recursion. Traces of a process's execution take the form of traces of observable events, which provide the basis for process calculi to reason about properties of a process's computation. Models such as CSP abstract away non-synchronizing concurrency by interleaving point-like events an observer perceives to have occurred simultaneously. There are several points to make here, the most important of which is a loss of entropy, which is described in detail by Smith (2000). Imposing an order of events where there was none, or where events overlap to some degree leads to a loss of information and limits what a process calculus may reason about. Denning's choice uncertainty principle motivates us to develop process algebras and calculi that permit reasoning directly about near-simultaneous events. EAP, in addition to being a model of true concurrency, introduces intermediate units of composition, actions, which are partial traces. This implies actions don't correspond to entire processes, but also means actions are a new way to reason about infinite processes, or more significantly, threads of execution, which cross process boundaries.

EAP provides a three-level process algebra for the modeling, study, and specification of concurrent processes. It should prove especially useful for modeling and reasoning about properties of multicore and modern grid computation (cf., for example, (Foster and Kesselman, 1998; Foster, 2002)). Operations in the EAP process algebra are defined in sections 3.3-3.5. There are unary operations, binary operations and operations of higher arity. Operations on the second and third levels of the EAP process algebra, i.e., operations of which operands are actions and processes, are also called operators. As we show in Section 4, EAP process algebra operations can faithfully represent operators from many conventional process algebras, such as CSP (Hoare, 1985; Reed and Roscoe, 1988), ACP (Bergstra and Klop, 1984; Baeten and Bergstra, 1991), TCSP (Brooks, et al, 1984), and CCS (Milner, 1980; Moller and Tofts, 1990).

The process algebra developed in this paper is based on a model of observation-based composition of processes, actions, and events. The set of observable events is a way from the process algebra to a process calculus, the next step in the development of EAP. The motivation for view-centric reasoning (VCR) stemmed from a desire to preserve more information about the history of a computation by introducing lazy observation, and multiple, possibly imperfect observers (Smith, 2000).

At the same time, compositions of systems and processes become an important tool for the development of modern hardware and software systems with complex sets of requirements. Unfortunately, as Neumann (2006) writes, there is a huge gap between theory and common practice: system compositions at present are typically ad hoc, based on the intersection of potentially incompatible component properties, and dependent on untrustworthy components that were not designed for interoperability–often resulting in unexpected results and risks.

Results of this paper provide additional evidence to support this statement. The analysis of real systems and processes made possible the ability to find new types for even such a well-known operation as sequential composition. This was possible because the authors considered not only data-flow relations between systems and processes used in the conventional sequential composition, but also temporal and causal relations.

In Section 2, we at first formulate general axioms for models of concurrency and introduce a new classification of these models. We introduce informal axioms to preserve connections with real processes. Such informal axioms were used by Euclid and are used now in theoretical physics. Only then we build the static stratum of the EAP metamodel as a unifying framework for concurrency studies. Its constructions use advanced means of modern mathematics: fiber bundles, manifolds, general algebraic systems, universal algebras and logical models. We develop the initial model in this generality to form a base for further development of this model because it is necessary to be able to represent in a model of concurrency not only properties of the system time and space, but also properties of the physical time and space. At the same time, contemporary physics uses these advanced mathematical structures to model physical objects and processes, including the physical time and space (cf., for example, (Bleecker, 1981)). Besides, some contemporary models of concurrency already use categories (cf., for example, (Pratt, 2003)) and algebraic topology (cf., for example, (Gunawardena, 1994)).

However, it is possible to take a simplified form of this model that utilizes mathematics no more sophisticated than basic set theory. In the majority of cases, it is sufficient to employ only finite sets or, at most, countable sets. However, to properly treat complex problems of concurrent systems and computations, such as time relativity, we need sufficiently powerful mathematical tools, which we can find only in advanced mathematics.

In Section 3, we, at first, develop algebraic foundations for modeling dynamics of multicomponent systems and concurrent processes and then build the dynamic stratum of the EAP metamodel. Here we base our models on structures and tools from universal algebra and topology. There is a dual approach to concurrency in the framework of category theory (cf., for example, (Pratt, 2003). It is possible to develop the EAP metamodel in the category setting. However, the categorical formalization of the EAP metamodel is studied elsewhere. Composition of concurrent processes is based on compositions of actions that constitute these processes, and composition of actions is based on compositions of events that constitute these actions. That is why the topic of the third section is composition of events and actions in order not to go beyond any reasonable length of the paper. Compositions of processes are constructed and studied elsewhere. Here we only show how processes evolve from actions and consider some simple properties of these processes.

In Section 4, relations between the EAP metamodel and other models of concurrent processes are considered. It is demonstrated that the EAP metamodel encompasses a variety of other models of distributed systems and concurrent processes.

For simplicity, we consider mostly point-like events in this paper and do not explicate to the full extent topological and temporal structures in systems of events, actions and processes. However, the EAP metamodel allows one to achieve a much deeper understanding of these relations and structures, and to use this model for organization of concurrent processes in distributed systems.

## 2. Models of Events, Actions, and Processes

Models of concurrency are classified with respect to two parameters: time and substance. Traditionally (cf., for example, (Sassone, et al, 1996)), eight types of models are separated based on three dimensions and six types: behavior/system, interleaving/noninterleaving, and linear/branching.

The first dimension is based on substance representation of processes, while two other dimensions are based on temporal aspects. Namely, interleaving means a possibility of linearization of the time scale, while linear model means utilization of a linear time scale and branching model means utilization of a tree-like (branching) time scale.

There are two substance types of process models and thus, of concurrency models: embodied (or system) and pure (or behavior) process (concurrency) models.

**Definition 2.1.** An *embodied process* (*concurrency*) *model* includes a model of the systems in which processes go on.

Petri nets (Petri, 1962; Peterson, 1991), synchronization trees (Winskel, 1985), labeled transition systems (Sassone, *et al*, 1996), and grid automata (Burgin, 2003; 2005) are examples of embodied process (concurrency) models.

**Definition 2.2.** In a *pure process* (*concurrency*) *model* only processes are represented in more or less detail.

The CSP model of Hoare (1985), CCS model of Milner (1989), ESP model of Lee and Sangiovanni-Vincentelli (1998), VCR model of Smith, (2000), and EVCR and EAP models of Burgin and Smith (2006; 2007) are examples of pure process (concurrency) models. Partial recursive and recursive functions are examples of pure computational models.

In turn, embodied process (concurrency) models can be divided into two classes: black box models and structural system models.

**Definition 2.3.** A *black box embodied process* (*concurrency*) *model* does not describe the structure of the system in which processes go on.

Synchronization trees (Winskel, 1985) and labeled transition systems (Sassone et al, 1996) are examples of black box embodied process (concurrency) models.

Finite automata and abstract state machines (Gurevich and Spielmann, 1997) are examples of black box embodied computational models.

**Definition 2.4.** A *structural system* (*white box*) *embodied process* (*concurrency*) *model* describes the structure of the system in which processes go on.

Petri nets (Petri, 1962; Peterson, 1991) and grid automata (Burgin, 2003; 2005) are examples of white box embodied process (concurrency) models.

Turing machines and inductive Turing machines are examples of white box embodied computational models (Burgin, 2005).

## 2.1. Events

The basic elements and concepts of a theory of concurrent processes are events. To build a mathematical model of a system of events, we, at first, formulate the main principles in the form of informal axioms that reflect fundamental properties of events. Then we build actions from events and processes from actions, introduce operations with events, actions, and processes, and formalize the obtained structures as algebras.

Note that our axioms for events are informal because informal axioms better reflect features of real events – formalization, as a rule, ignores some properties and relations. Formal axioms are used for models of events in process algebra, considered as an algebraic theory to formalize and study by mathematical tools systems of concurrent processes.

**Axiom E1.** Events exist in space and time and have some meaning.

Space and time can be physical or system. Physical time means the time in a physical region (physical space) where the event takes place. System time means the inner time of the system in which the event takes place (Fujimoto, 1997; Boukerche, Das, Datta, and LeMaster, 2001). Such a system can be a separate computer, a local network of computers or a global network, such as the Internet. According to the system theory of time, system time can be essentially different from physical time (Burgin, 2002). For instance, when we consider a theoretical model of computation, such as a Turing machine, all its operations have the same duration equal to one.

At the same time, we know that computer operations can have essentially different duration. For example, example multiplication of numbers takes more time than addition of numbers, multiplication with the floating point takes more time than multiplication with the fixed point, and multiplication of matrices takes more time than multiplication of numbers.

Both kinds of time, system and physical, are used in temporal databases (cf., for example, (Snodgrass and Jensen, 1999; Date, et al, 2002)). Namely, the temporal aspects usually include valid time and transaction time. These attributes go together to form bitemporal data.

*Valid time* is the physical time and is used to denote the time period during which a fact is true with respect to the real world.

*Transaction time* is the system time of the database and is used to form the time period during which a fact is stored in the database

Physical space is the space where people live, physical processes go on, and physical (or material) systems exist. Physical space is modeled in different physical theories.

System space depends on the representation of the system(s) where events take place. For instance, the system space for such systems as finite automata or labeled transition systems (Sassone et al, 1996) is the state space of a finite automaton or labeled transition system. As we take such a system as a cellular automaton $A$, then the system space of events in $A$ is the corresponding grid of state spaces of individual finite automata that constitute $A$ (Codd, 1968). In a Petri net, the system space is the network in which places and transitions are situated (Petri, 1962; Peterson, 1981).

The meaning of an event is a description of what happened in this event. For instance, the meaning of one event in a computer can be "sending an e-mail", the meaning of another event in a computer can be "reading from a CD", and the meaning of an event in a printer can be "printing".

To develop semantics for events, we consider three types of events: abstract, process and embodied or system events.

Considered at the beginning of this section, types of process models utilize five types of events: abstract or pure events, pure temporal events, and three types of embodied or system events – pure spatial events, white-box embodied events and black-box embodied events.

**Definition 2.5.** An *abstract* or *pure event* is a representation of what happened (of what is done), but not an indication of when or where.

In some sense, abstract events are not events but types of events or event variables. Another kind of event variable is an event in which the event's name or meaning is variable.

**Definition 2.6.** A *pure temporal event* is a representation of what happened, indicating a time of occurrence, but not where it happened.

The majority of concurrency models that include time, such as timed CCP (Nielsen and Valencia, 2004), use only pure temporal events or actions.

**Definition 2.7.** An *embodied* or *system event* is a representation of what happened, indicating both a time and place of happening.

The most popular concurrency model that uses embodied/system events is a Petri net.

**Definition 2.8.** A *black box embodied event* is a representation of what happened without a description of the system structure.

Synchronization trees (Winskel, 1985) and labeled transition systems (Sassone, et al, 1996) are examples of black box embodied process (concurrency) models.

Finite automata and abstract state machines (Gurevich and Spielmann, 1997) are examples of black box embodied computational models.

**Definition 2.9.** A *white box* (*structural*) *embodied event* is a representation of what happened without a description of the system structure.

Usually, a system belongs to some environment, while a process interacts (communicates) with other processes, which collectively form the environment of the first process. Taking into account this natural structure, we come to three categories of events that have different meanings.

**Axiom E2a.** The meaning of an *event in a system* or in *a process* is a change in this system or in this process.

Axiom E2a interprets events (in a system, or of a process) as changes, or as reflections of changes, [of the state or phase] of the system (in the process). For instance, an event in such a system as a finite automaton or finite state machine is a state transition of this automaton or machine. Sending a message is an event in a network. Printing a symbol or a text, producing a symbol on the screen of the display, and calculating 2 + 2 are events in a computer. The latter event can be also an event in a calculator.

**Axiom E2b.** The meaning of an *event by a system* is a change in the environment caused by this system or process.

For instance, when a system $R$ sends information to another system or builds a new system or takes something from one place to another, it is an event by this system $R$.

**Axiom E2c.** The meaning of an *event for a system* is a change in relations between this system or process and its environment.

For instance, when a system $R$ was connected to another system, and then this connection is broken, it is an event for this system $R$.

Note that even in the best embodied models of concurrency, such as Petri nets (Petri, 1962; Peterson, 1991) or labeled transition systems (Sassone et al, 1996), semantics represents only events in a system, while in reality two other types of events are of no less importance.

To build an efficient and flexible model of concurrent processes, developing algebras and calculi for them, we need mathematical structures that allow one to deal with a diversity of situations that emerge in practice.

Recall the definitions of some mathematical concepts that we need in our model.

An *algebraic system* is a set with relations (predicates) and operations in this set (Malcev, 1970).

When there are only operations, we have algebras, or universal algebras (Cohn, 1965). When there are only relations or predicates, we have mathematical structures or models when these structures are related to corresponding logical theories (Chen and Keisler, 1966; Robinson, 1963). In mathematics, usually two types of relations are considered relations of order, including preorder, tolerance and equivalence, and topological relations (Bourbaki, 1960).

A *topological manifold* is a Hausdorff topological space $X$ in which every point has a neighborhood homeomorphic to an open subset of Euclidean space $\mathbf{R}^n$. The integer $n$, called the *dimension* of $X$, must be the same for all points of $X$.

A *fiber bundle* $\mathbf{B}$ (also called fibre bundle) is a triad $(E, p, B)$ where the topological space $E$ is called the *total space* or simply, *space* of the fiber bundle; the topological space $B$ the *base space* or simply, *base* of the fiber bundle; and $p$ is a topological projection of $E$ onto $B$ such that every point in the base space has a neighborhood $U$ such that $p^{-1}(b) = F$ for all points $b$ from $B$ and $p^{-1}(U)$ is homeomorphic to the direct product $U \times F$. The topological space $F$ is called the *fiber* of the fiber bundle $\mathbf{B}$. Informally, a fiber bundle is a topological space which looks locally like a product space $U \times F$.

It is possible to find introductory information on topological manifolds, fiber bundles and other topological structures in (Gauld, 1974; Lee, 2000; Steenrod, 1951; Husemoller, 1994)).

A *multiset* (sometimes called a *bag*) as a set consists of elements, but in contrast to sets each member of a multiset has a *multiplicity*, which indicates *how many* times it is present in this multiset (cf., for example, (Knuth, 1997)). For instance, in the multiset { $a$, $a$, $b$, $b$, $b$, $c$ }, the multiplicities of the members $a$, $b$, and $c$ are respectively 2, 3, and 1.

To formally represent temporal, spatial and semantic characteristics of events, we use three algebraic systems: a naming system $N$ of names, a semantic system $V$ of values, and a space-time system $ST$ of tags.

In the simplest case, *N*, *V* and *ST* are sets as in the ASP model (Lee and Sangiovanni-Vincentelli, 1996). Usually, systems *V* and *ST* are topological manifolds with some algebraic operations and relations, i.e., they are algebraic systems. For example, physical space-time is a four-dimensional manifold in relativity theory, ten-dimensional manifold in string theory and eleven-dimensional manifold in superstring theory (cf., for example, (Green, 2000; Zwiebach, 2004). Important relations for events are temporal relations, in particular, order relations, introduced and studied by the authors in (Burgin and Smith, 2006a). The naming system *N* is usually a multiset (Knuth, 1997). Sometimes the naming system *N* has some operations in it.

The *naming system N* consists of names of events. A name shows the type of an event. Names can be organized in a hierarchical system.

The *semantic system V* consists of values of events. A *value v* describes (indicates) what happened (what is done), determining a transformational or operational semantics of events (cf., for example, (Plotkin, 1981; Gunter, 1992; Winskel, 1993)).

There is a function (correspondence, in a general case) *sem*: $N \to V$ from *N* to *V* that gives meaning to events, as well as to types of events.

The *space-time* (or *tag*) system *ST* consists of tags of events. A *tag t* shows where and when something happened (is done). It consists of time and place, determining location of events. Topologically the space-time manifold *ST* is a fiber bundle ***ST*** with the base *T* and the fiber *S* (cf. Fig. 1).

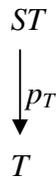

**Fig. 1**. The space-time fiber bundle

It means that $ST = (ST, p_T, T)$ where $p_T$ is a continuous projection and $p_T^{-1}(t) = S$ for any point $t$ from $T$. Here $T$ is the time manifold, which is naturally decomposed into physical time scale $T_{ph}$ and system time scale $T_{st}$ (cf. Fig. 2).

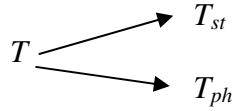

**Fig. 2.** Time manifold decomposition

Usually time in computer science is linear, implicit and discrete. However, studies of concurrency brought researchers to consider with more complex scales to be able to correctly reflect computational reality. For instance, branching time plays an important role in many computational models of concurrency (cf., for example, (de Bakker, 1989; Emerson and Halpern, 1986)). Transitions of higher dimensional automata (cf., for example, (Pratt, 1991; Goubault, 1993)) go in a multidimensional time.

The fiber $S$ is the space manifold. It is naturally decomposed into physical space model $S_{ph}$ and system space $S_{st}$ (cf. Fig. 3).

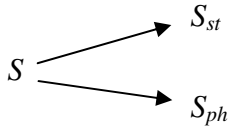

**Fig. 3**. Space manifold decomposition

For simplicity, we take $T_{ph} = \bigcup_{a \in S_{st}} T_{pha}$ where $a$ is a node in the considered system, $T_{pha}$ is the time scale of the physical place where $a$ is situated, and all $T_{pha}$ are linearly ordered. To make the time scale more transparent, we assume that each $T_{pha}$ is a subset of the set $R$ of all real numbers.

In a similar way, $T_{st} = \bigcup_{a \in S_{st}} T_{sta}$ where $T_{sta}$ is the time scale of time at a node $a$. Time with the scale $T_{sta}$ corresponds to the concept of local time used in some theories of concurrency (cf., for example, (Mazurkiewicz, 1977; Pratt, 1992)).

In this context, an event $e$ is equal to a pair $(d, t)$ where $d$ is the name of $e$ and belongs to $N$ and $t$ is the tag of $e$. For a *point-like* event $e$, its tag $t$ belongs to $ST$. There are two natural mappings of events onto their names and tags, namely, if $e = (d, t)$, then $\mathbf{n}(e) = d$ and $\mathbf{t}(e) = t$. The mapping $\mathbf{n}$ allows us to define meaning (semantics) of events as Sem $e = \text{sem}(\mathbf{n}(e))$. In general, we assume that the same event may have different names, i.e., it is possible that $e = (d, t) = (a, t)$.

Note that any component of an event, the name, value, tag, time coordinate and/or space coordinate, can be a variable. In this case, i.e., when, at least, one component of an event $e$ is a variable, the event $e$ is potential.

By $\mathbf{EV}_{NV, ST}$ we denote all (point-like) events with names from $N$, semantics from $V$ and tags from $ST$. For simplicity, the set $ST$ is considered as the direct product of the time scale $T$ and the abstract space $S$. There are two canonical projections: $p_S : ST \rightarrow S$ and $p_T : ST \rightarrow T$. Thus, time of an event $e$ is $p_T(\mathbf{t}(e))$ and the place of an event $e$ is $p_S(\mathbf{t}(e))$. In the simplest case, a tag $t$ of a point-like event is a pair $(p_S(t), p_T(t))$.

In computers and networks, the time scale is usually an infinite discrete set of points isomorphic to the set $N$ of all natural numbers. The space of a network usually is some graph (cf., for example, (Burgin, 2005)).

In addition to point-like events, there are other types of events.

In *extended events*, time (that is, the projection of the tag $t$ on the time manifold) is an interval, which is represented by two points: the beginning and the end of this interval. For instance, if we assume that $ST$ coincides with $T$ and $T$ is the scale of the physical time, then an extended event $e = (d, (t_b, t_e))$, where $t_b$ is the time of the beginning, and $t_e$ is the time of the end, of the event $e$. Many models of concurrent processes ignore the space coordinate of events and assume that events take place in the physical time (cf., for example, (Lee and Sangiovanni-Vincentelli, 1996)).

If we consider extended events with tags that represent both time and space, then the space of the event can be not a single coordinate of the space manifold *S*, but a region in (subset of) this manifold. Thus, the possible values of extended event tags are pairs of regions (subsets): one in *S* and another in *T*.

Extended events can be separated into three classes: *spatial extended events*, *temporal extended events*, and *full-scale extended events*.

A *temporal extension* of an event means that the temporal component of the event tag is a name of some region (usually, an interval when the time scale is one-dimensional) in the time manifold (time scale) *T*. Such extended events are studied in (Burgin and Smith, 2006; 2006a),

A *spatial extension* of an event means that the spatial component of the event tag is a name of some region (e.g., a sphere or ball) in the space manifold *S*. For instance, if we assume that *ST* coincides with *S*, i.e., time is not considered, then it is natural to consider balls as values for tags because, as a rule, the space where an event occurs is not a point. In this case, an extended event may have the form $e = (d, (x, r))$ where *x* is the coordinates of the center of a sphere in which this event does on and *r* is the radius of this sphere. An example of a system in which time of events is ignored is a finite automaton. In full-scale extended events, tags represent regions in the space manifold *S* and in the time manifold *T*.

By $\mathbf{EV}_{NV,\,EST}$ we denote all extended events with names from *N*, semantics from *V*, and tags from P(*ST*) where is the set of all subsets from *ST*.

There are also *pointed events*, which have the form $e = (d, (t_1, t_2, \ldots, t_k))$. A possible interpretation of pointed events is given by the situation when an event $e = (d, (t_1, t_2, \ldots, t_k))$ is considered in different time scales and $t_i$ is the point-like time of *e* in the scale $T_i$ ($i = 1, 2, 3, \ldots, k$). These different scales $T_i$ may reflect physical time or system time at different places (the so-called, local time (Mazurkiewicz, 1977; Pratt, 1992)) or represent system time (cf. Burgin, 2002) of those distributed systems in which this event takes place. Another interpretation of pointed events is given by the situation when an event $e = (d, (t_1, t_2, \ldots, t_k))$ represents an abstract event that happens at several places.

By $\mathbf{EV}_{NV,\,kST}$ we denote all pointed events with names from $N$, semantics from $V$, and tags from $ST^k$ where is the set of all subsets from $ST$.

Assuming that spaces $V$ and $ST$ give a complete representation of all events, we have the following axiom.

**Axiom E3.** An event is uniquely defined by time, space and what happened (what is done).

In the formal model, it means that if $e_1 = (d_1, t_1)$, $e_2 = (d_2, t_2)$, $v_1 = v_2$, $sem\ d_1 = sem\ d_2$, and $t_1 = t_2$, then $e_1 = e_2$.

Axiom E3 implies that embodied or system events form a set. Abstract or pure events form, as a rule, multisets because the same anbstract event can occur several times at different places or at different time.

When spaces $V$ and $ST$ do not give a complete representation of all events, it is also natural to consider systems of events as multisets (cf., for example, (Smith, 2000)). For instance, often an event is identified with its value $v$, while time and place are ignored. In this case, a process can contain several copies of the same event. At the same time, it is possible that two or more processes contain different copies of an event in some cases, while they have a common copy (or several common copies) of the same event. For instance, taking such an event as sending a message, we see that different copies of this event belong, as a rule, to many different processes. At the same, each copy of such an event as information (data) exchange between two processes necessarily belongs to two processes. Parallel events can coincide as abstract events.

A similar situation emerges when the space (set) $ST$ does not give a complete representation of all events. For instance, the space of the event is not taken into consideration or the time of the event is taken as a point instead of an interval.

To study information processing (computation, communication, etc.), we consider events of two types: changes of information carriers (symbols, signals, data, knowledge, etc.) and changes of information processing systems.

**Example 2.1.** If we model information processing in such a system as a Turing machine, then events of the first type are writing and erasing symbols in cells of the machine memory. Events of the second type are movements of the head of the Turing machine.

**Definition 2.10.** a) Two point-like events $e_1 = (d_1, t_1)$ and $e_2 = (d_2, t_2)$ are called *parallel* or *simultaneous* if $p_T(t_1) = p_T(t_2)$.

b) Two extended events $e_1 = (d_1, ((t_{1b}, t_{1e}), s_1))$ and $e_2 = (d_2, ((t_{2b}, t_{2e}), s_2))$ are called *parallel* if $t_{1b} = t_{2b}$ and $t_{1e} = t_{2e}$.

We denote this by $e_1 \parallel e_2$. Informally, this means that two events are parallel when they occur exactly at the same time. This agrees with one's general intuition and is represented on the abstract level by the simultaneity relation, $ST_R$, previously introduced by the authors in (Burgin and Smith, 2006; 2006a).

**Example 2.2.** Let us consider information processing in such a system as a Turing machine. In the traditional representation (cf., for example Rogers, 1987), each step of a Turing machine can include two parallel operations: changing the state of the machine and either writing in a cell or erasing from a cell, or moving the head of the machine. Contemporary textbooks (cf., for example (Hopcroft, Motwani and Ullman, 2001)) allow the Turing machine to do three parallel operations in one step.

Note that what is parallel and even simultaneous events depends on the chosen time scale. For instance, switching in multithreading can happen so fast as to give the illusion of simultaneity to an end user, that is, for the computer events before and after switching look simultaneous, while to the computer where these events happen they are not simultaneous.

**Proposition 2.1.** Parallelism of point-like (extended) events is an equivalence relation.

<u>Proof</u>. Let us consider, at first, point-like events $e_1 = (d_1, t_1)$ and $e_2 = (d_2, t_2)$. By Definition 2.10, any point-like event is parallel to itself. It means that parallelism of point-like events is a reflexive binary relation. Besides, if $e_1 \parallel e_2$, then $p_T(t_1) = p_T(t_2)$. Thus, by Definition 2.10, $e_2 \parallel e_1$, i.e., parallelism of point-like events is a symmetric binary relation. Finally, if $e_1 \parallel e_2$ and $e_2 \parallel e_3$, then $p_T(t_1) = p_T(t_2)$ and $p_T(t_2) = p_T(t_3)$. Consequently, $p_T(t_1) = p_T(t_3)$ and by Definition 2.10, $e_1 \parallel e_3$. Thus, parallelism of point-like events is a transitive binary relation, and by definition, it is an equivalence relation.

Similar reasoning gives us a proof that parallelism of extended events is an equivalence relation.

Proposition 2.1 is proved.

However, in practice, it is impossible to make an unambiguous choice between near-simultaneous events under a deadline (Denning, 2007). Thus, we need to introduce approximately or fuzzy parallel events. To do this, we use an arbitrary positive real number $r$ as the parameter of this approximation or fuzziness.

**Definition 2.11.** a) Two point-like events $e_1 = (d_1, t_1)$ and $e_2 = (d_2, t_2)$ are called *r-parallel* or *r-simultaneous* if $|p_T(t_1) - p_T(t_2)| < r$.

b) Two extended events $e_1 = (d_1, ((t_{1b}, t_{1e}), s_1))$ and $e_2 = (d_2, ((t_{2b}, t_{2e}), s_2))$ are called *r-parallel* if $|t_{1b} - t_{2b}| < r$ and $t_{1e} = t_{2e} < r$.

We denote this by $e_1 \parallel_r e_2$. Informally, this means that two events are approximately or fuzzy parallel when the difference between times of their occurrence is less than some number $r$ that represents indistinguishability of event time coordinates.

For representing communication processes, we need events of four types (Burgin, 1997): emission, reception, reading (or information extraction), and writing (or information insertion). We introduce the following corresponding set notations:

**Em**$_{NV, ST}$ denotes the set of all emissions from **EV**$_{NV, ST}$.

**Em**$_{NV, EST}$ denotes the set of all emissions from **EV**$_{NV, EST}$.

**Em**$_{NV, kST}$ denotes the set of all emissions from **EV**$_{NV, kST}$.

**Rc**$_{NV, ST}$ denotes the set of all receptions from **EV**$_{NV, ST}$.

**Rc**$_{NV, EST}$ denotes the set of all receptions from **EV**$_{NV, EST}$.

**Rc**$_{NV, kST}$ denotes the set of all receptions from **EV**$_{NV, kST}$.

**Rd**$_{NV, ST}$ denotes the set of all readings from **EV**$_{NV, ST}$.

**Rd**$_{NV, EST}$ denotes the set of all readings from **EV**$_{NV, EST}$.

**Rd**$_{NV, kST}$ denotes the set of all readings from **EV**$_{NV, kST}$.

**Wr**$_{NV, ST}$ denotes the set of all writings from **EV**$_{NV, ST}$.

**Wr**$_{NV, EST}$ denotes the set of all writings from **EV**$_{NV, EST}$.

**Wr**$_{NV, kST}$ denotes the set of all writings from **EV**$_{NV, kST}$.

All these events are communication events. A general definition of a communication event is based on the classification of information operations in the general theory of information (Burgin, 1997).

**Definition 2.12.** An event $e = (d, t)$ is called a communication event if *sem d* is a communication operation.

Communication operations are changes of information carriers for portions of information (Burgin, 1997). For instance, readings some text by a word processor from a CD changes the CD as the information carrier for this text to the screen of the computer as the information carrier for this text.

Receiving some picture from a channel, e.g., a cable, by a TV set is a communication event in which the channel as the information carrier for this picture is changed to the screen of the TV set as the information carrier for this picture.

## 2.2. Actions

Usually processes are formed from events (cf., for example, (Hoare, 1985 and Milner, 1989)). However, here, following (Lee and Sangiovanni-Vincentelli, 1996), we use one more level of hierarchy that consists of actions. Three levels of hierarchy reduce complexity of descriptions and make the model more flexible.

**Definition 2.13.** An *action A* is a system of related events, i.e., $A = (e_1, e_2, \ldots, e_n)$. The number $n$ is called the *cardinality* of $A$ and denoted by $|A|$.

It is necessary to understand that the action $A$ is defined neither as a set of events $e_1, e_2, \ldots, e_n$, nor as a multiset of events $e_1, e_2, \ldots, e_n$, nor as a vector or $n$-tuple of events $e_1, e_2, \ldots, e_n$. A system is defined not only by its elements, but also by relations between these elements (Mesarovic and Takahara, 1975). Thus, what is $A$ depends not only on events $e_1, e_2, \ldots, e_n$, but also on relations between these events. If there are no such relations in $A$, then $A$ is a set of events. If there are only relations of indiscernibility of events in $A$, then $A$ is a multiset of events. If there is a relation of linear order in $A$, then $A$ is an ordered $n$-tuple of events. If there is a relation that connects some events in $A$, then $A$ is a graph (network) with events as vertices.

Note that in some models of concurrency, such as ACP (Bergstra and Klop, 1984), actions are indivisible primitives of the model and processes are built from actions. To represent such primitive actions in EAP, we consider actions such that each consists of a single indivisible (primitive) event.

There are different interpretations of actions. In one of them, actions in the EAP are theoretical representations of threads and fibers. Threads are a way for a program to split itself into two or more simultaneously (or pseudo-simultaneously) running tasks. Fibers realize the same function as threads. The difference is that fibers use cooperative multitasking, while threads use pre-emptive multitasking. Threads and processes differ from one operating system to another, but in general, how a thread is created and shares its resources is different from the way a process does.

Multiple threads can be executed in parallel on many computer systems. This process is called multithreading. When only one processor is used, multithreading is generally organized by time slicing (similar to time-division multiplexing), wherein a single processor switches between different threads, in which case the processing is not literally simultaneous, for the single processor is really doing only one thing at a time. It is an example of interleaving of actions.

Threads form a different level from traditional multitasking operating system processes because processes are typically independent, carry considerable state information, have separate address spaces, and interact only through system-provided inter-process communication mechanisms. Threads, on the other hand, typically share the state information of a single process, and share memory and other resources directly. Threads are parts of processes. At least one thread exists within each process. If multiple threads can exist within a process, then they share the same memory and file resources. Threads are pre-emptively multitasked if the operating system's process scheduler is pre-emptive. This allows the operating system to do switching between threads in the same process faster than switching between processes. For instance, systems like Windows NT and OS/2 are said to have "cheap" threads and "expensive" processes.

Another interpretation is a signal from the metamodel ESP of Lee and Sangiovanni-Vincentelli (1996). For simplicity, we consider only point-like events in what follows. There are special classes of actions.

**Definition 2.14.** An action *A* is called *finite* if it consists of a finite number of events. Otherwise, *A* is called *infinite*.

**Definition 2.15.** An action *A* is called a *signal* if it consists of communication events.

For instance, writing or erasing a symbol in a cell of the Turing machine memory is a signal. Movements of the head of a Turing machine are not signals, but pure actions. Changing the place of an information carrier, e.g., sending a letter by regular mail, is also a signal.

Existing tradition resulted in a situation where communication and computation processes are mostly studied separately. In reality, any computation includes some communication operations and any communication includes some computation operations. Here we put an emphasis on communication of processes but build our model so that it allows one to represent all types of processes and operations used in practice.

**Definition 2.16.** An action $A$ is called an *explicit communication action* if it has at least one emission and/or reception.

In an explicit communication action $A$, it is possible to separate three groups of events: emissions, receptions, and all other events. As a result, we represent $A$ in the form of a triad $A = (A_{rc}, A_{gen}, A_{em})$ where $A_{rc}$ is the set of all receptions from $A$, $A_{em}$ is the set of all emissions from $A$, and $A_{gen}$ is the set of all other events from $A$.

**Definition 2.17.** An action $A$ is called an *implicit communication action* if it has at least one event of reading from, and/or writing to, a shared information carrier.

For instance, two processes (or programs) have a common memory. Then writing to and reading from this memory allows these processes to communicate with one another. Such memory is a specific instance of a *communication space* introduced in (Burgin, Liu and Karplus, 2001). In general, a communication, or interaction, space is a special media often used for human-computer interaction. Another example of such a memory is the concept of a *store* used in concurrent constraint programming (Nielsen and Valencia, 2004; Saraswat, Rinard, and Panangaden, 1991; Valencia, 2005).

**Definition 2.18.** An explicit communication action $A$ is called a *pure explicit communication action* if it contains only emissions or/and receptions.

Thus, a pure explicit communication action $A$ has the form of a dyad $A = (A_{rc}, A_{em})$.

**Definition 2.19.** An action $A$ is called an *action without repetition* if all events from $A$ have different values, i.e., if $e_i = (d_i, t_i)$ and $e_j = (d_j, t_j)$ are events from $A$, then Sem $d_i \neq$ Sem $d_j$.

Informally, this means that nothing happens twice in an action without repetition.

**Definition 2.20.** An action $A$ is called a *coordinated action* if all events from $A$ have different tags, i.e., if $e_i = (d_i, t_i)$ and $e_j = (d_j, t_j)$ are events from $A$, then $t_i = t_j$ implies $e_i = e_j$.

Informally, this means that, at the same place and same time, only one thing can happen (be done) in a coordinated action.

**Definition 2.21.** An action $A$ with point-like events is called *sequential* if all events from $A$ have different times, i.e., if $e_i = (d_i, t_i)$ and $e_j = (d_j, t_j)$ are events from $A$, then $p_T(t_i) = p_T(t_j)$ implies $e_i = e_j$.

Informally, this means that, at the same time, only one thing can happen (be done) in a sequential action. In other words, a sequential action does not have parallel events.

The cardinality of a sequential action $A$ is also called the *length* of $A$.

**Proposition 2.2.** Any sequential action $A$ is coordinated.

It is necessary to discern action that were realized, i.e., actions that consist of events that already happened, and potential actions, which represent alternate versions of what may happen. This gives us three types of actions:

- An *actualized action* is an action, for which a realization of this action includes all its events.
- A *potential action* is an action, for which includes all events for all possible realizations of this action.
- An *emerging action* is an action that consists of two parts (subactions): one is an actualized action, while another is a potential action. When we take with the conventional linear scale, such as physical time in Newtonian mechanics, then all events from the first (actualized) part of an emerging action must precede all events from the second (potential) part of this action.

For instance, the choice operation +, which is used in many models of concurrency for processes (cf., for example, ((Bergstra and Klop, 1984)), but can be also applied to actions, transforms any two actions into a potential action.

Potential actions are schemas (Burgin, 2006), plans or programs for realization. Any realization of a potential action is an actualized action.

Possible realizations of a potential action $A$ are determined by the following compatibility relations in $A$.

Two events $e_i$ and $e_j$ from $A$, are *compatible* if any realization of $A$ that contains $e_i$ also contains $e_j$.

Two events $e_i$ and $e_j$ from $A$, are *weakly compatible* if any realization of $A$ that contains $e_i$ can also contain $e_j$.

Two events $e_i$ and $e_j$ from $A$, are *strongly compatible* if any realization of $A$ that contains $e_i$ must also contain $e_j$.

Two events $e_i$ and $e_j$ from $A$, are *incompatible* if any realization of $A$ that contains $e_i$ does not contain $e_j$.

Two events $e_i$ and $e_j$ from $A$, are *strongly incompatible* if any realization of $A$ that contains $e_i$ cannot contain $e_j$.

Two events $e_i$ and $e_j$ from $A$, are *weakly incompatible* if a realization of $A$ that contains $e_i$ may not contain $e_j$.

Let us consider an action $A$ and a set $R$ of relations in $A$, i.e., relations between events from $A$ that determine $A$ as a system.

**Definition 2.22.** a) Any subset $B$ of an action $A$ in which all relations between events in $B$ are induced by relations from $R$ are preserved is called an *R-subaction* of $A$.

b) An $R$-subaction of $A$ is called a *complete subaction* of $A$ if $R$ consists of all relations in $A$.

When $R$ is not specified, we call $B$ a *subaction* of $A$.

These definitions imply that the cardinality of any proper subaction of an action $A$ is less than the cardinality of $A$.

**Proposition 2.3.** Any subaction of a finite action $A$ is a finite action.

**Proposition 2.4.** Any subaction of an actualized action $A$ is an actualized action.

For potential and emerging actions, this result is not true in a general case.

**Proposition 2.5.** Any $R$-subaction of a signal (communication action, pure communication action) is a signal (correspondingly, communication action, pure communication action).

**Theorem 2.1.** A complete subaction of a coordinated action (action without repetition) $A$ is a coordinated action (action without repetition).

**Proposition 2.6.** A (complete) subaction $C$ of a (complete) subaction $B$ of an action $A$ is a (complete) subaction of $A$.

To build parallel compositions of actions and processes, it necessary to have a relevant and sufficiently formalized definition of parallel actions and processes. However, the relation "to be parallel" is not so simple as the relation "to be sequential". As a rule, the relation "to be parallel" is a fuzzy or valued relation (property). That is why, we start with measures of parallelism.

Let us consider two actions $A = (e_{11}, e_{12}, \ldots, e_{1n})$ and $B = (e_{21}, e_{22}, \ldots, e_{2n})$.

**Definition 2.23.** The measure $m^{par}(A, B)$ of parallelism between $A$ and $B$ is equal to the number of pairs $e_{1i} \parallel e_{2j}$.

**Definition 2.24.** The measure $m^{par}(A)$ of parallelism in an action $A$ is equal to the number of parallel pairs $e_{1i} \parallel e_{1j}$ in the action $A$.

**Definition 2.25.** The measure $m_{par}(A)$ of parallelism in the action $A$ is equal to the difference between the cardinality of $A$ and the cardinality (length) of a maximal sequential subaction of $A$.

Note that, in a general case, a maximal sequential subaction of an action is not unique. So, to show that our definition is consistent, we need to demonstrate that any two maximal sequential subactions of an action $A$ have the same cardinality.

Indeed, let us consider two maximal sequential subactions $B$ and $C$ of an action $A$ and assume that the length of $B$ is larger than the length of $C$. Then $B$ has an event $e$ that is not parallel to any event in $C$ because $B$ as sequential sequence does not contain parallel events. Then if we add the event $e$ to the actions $B$, we will still have a sequential subaction of $A$ and this subaction properly contains $C$. This contradicts our assumption that $C$ is a maximal sequential subaction of $A$ and thus, proves that subactions $B$ and $C$ have the same cardinality (length).

**Lemma 2.1.** a) $m_{par}(A) \leq m^{par}(A)$ for any action $A$.

b) $m_{par}(A) = m^{par}(A) = 0$ if and only if $A$ is a sequential action.

c) $m_{par}(A) = m^{par}(A)$ if and only if either there are no parallel events in $A$ or any group of parallel events in $A$ contains only two elements.

**Definition 2.26.** The measure $m_{par}(A, B)$ of parallelism between $A$ and $B$ is equal to the number of parallel pairs of events in two maximal sequential subactions of $A$ and $B$.

In other words, if $D$ is a maximal sequential subaction of $A$ and $C$ is a maximal sequential subaction of $B$, then $m_{par}(A, B) = m^{par}(D, C)$.

**Lemma 2.2.** a) $m_{par}(A, B) \leq m^{par}(A, B)$ for any action $A$.

b) $m_{par}(A, B) = m^{par}(A, B)$ if and only if both $A$ and $B$ are sequential actions.

**Proposition 2.7.** If $D$ and $C$ are subactions of actions $A$ and $B$, correspondingly, then $m^{par}(D, C) \leq m^{par}(A, B)$ and $m_{par}(D, C) \leq m_{par}(A, B)$.

Let $0 \leq k \leq 1$.

**Definition 2.27.** An action $A$ is called *k-parallel* if $k \leq m_{par}(A)/|A|$.

**Definition 2.28.** Actions $A$ and $B$ are called *k-parallel* if $k \leq m_{par}(A, B)/\min\{|A|, |B|\}$.

Any action $A$ is 1-parallel to itself.

It is possible that each of two actions $A$ and $B$ is *k*-parallel to some action $C$, but actions $A$ and $B$ are not *k*-parallel to one another.

## 2.3. Processes

In our model, processes are built from actions as actions are built from events.

**Definition 2.29.** A *process P* is a system of actions, i.e., $P = (A_1, A_2, \ldots, A_m)$.

Systems are determined by their elements and relations between these elements (Mesarovic and Takahara, 1975). Thus, a process can define some additional relations between actions that constitute this process. Relations between actions in a process determine how actions are organized in this process. For instance, actions can form a sequence (a thread), a set without any relations, an arbitrary graph, and so on.

There are temporal relations between actions (cf., for example, (Burgin and Smith, 2006)) and inheritance relations when one action starts in a situation (state) that (may be, partially) resulted from the realization of another action. It is possible to define Temporal relations between actions can be induced by temporal relations between events that constitute these actions. However, temporal relations between events do not always completely determine temporal relations between actions. In this case, we need additional relations. These relations can be correspondences between time scales when events with tags that have values in time scales are used in actions. Additional temporal relations can also be precedence relations indicating when one of actions precedes another in the given process or how some events precedes other events in the actions that constitute the given process.

Similar to communication actions, we define communication processes.

**Definition 2.30.** A process *P* is called a *communication process* if it consists of communication actions.

**Definition 2.31.** A process *P* is called a *pure communication process* if it consists only of pure communication actions.

**Definition 2.32.** A process *P* is called *finite* if it consists of a finite number of finite actions. Otherwise, *P* is called *infinite*.

For a long time, computer scientists studied and accepted only finite processes. The finiteness condition was even included in definitions of the main mathematical models of algorithms, such as Turing machines. However, infinite processes become more and more important in theoretical studies and practical applications (cf., for example (Burgin, 2005; Park, 1980; Rabin, 1969; Vardi, 1994)).

**Definition 2.33.** Any subset of a process *P* in which all relations between events are induced by relations in *P* is called a *subprocess* of *P*.

**Proposition 2.8.** Any subprocess of a subprocess of *P* is itself a subprocess of *P*.

**Proposition 2.9.** Any subprocess of a pure communication or/and finite process *P* is itself a pure communication or/and finite process.

For communication processes, this result in general is not true.

**Definition 2.34.** Any subset *Q* of a process *P* in which relations between events include all relations induced by relations in *P* is called an *enhanced subprocess* of *P*.

Thus, any subprocess of a process is an enhanced subprocess of the same process.

**Proposition 2.10.** Any enhanced subprocess of an enhanced subprocess of *P* is itself an enhanced subprocess of *P*.

**Proposition 2.11.** Any enhanced subprocess of a pure communication or/and finite process *P* is itself a pure communication or/and finite process.

Similar to actions, it is necessary to discern processes that were realized, i.e., processes that consist of actions that already happened, and potential processes, which represent alternate versions of what may happen. This gives us three types of processes:

- An *actualized process* is a process that consists of actualized actions.
- A *potential process* is a process, for which includes all actions and events for all possible realizations of this process.
- An *emerging process* is a process that consists of two parts (subprocesses): one is an actualized process, while another is a potential process. When we take with the conventional linear scale, such as physical time in Newtonian mechanics, then all events from the first (actualized) part of an emerging process must precede all events from the second (potential) part of this process.

For instance, the choice operation +, which is used in many models of concurrency (cf., for example, ((Bergstra and Klop, 1984)), transforms any two processes into a potential process.

Potential processes are schemas (Burgin, 2006), plans or programs for realization. Any realization of a potential process is an actualized process. Note that it is possible that a potential process consists only of actualized actions. Examples of such processes give the choice operation applied to two actualized actions.

Temporal relations between events allow us to separate special classes of events.

**Definition 2.35** (Burgin and Smith, 2006; 2006a). *A coexisting pair of events* consists of two events where one starts before the next ends.

**Definition 2.36** (Burgin and Smith, 2006; 2006a). *A separable pair of events* consists of two events that are not coexisting.

To formalize the condition of action separability, we define the beginning and the end of an action from some action $A$. Namely, the beginning $t_{bA}$ of an action $A$ is equal to

$$t_{bA} = \min \{ p_T(t) ; e \in A, \text{ and } t \text{ is the tag of } e\}$$

The end $t_{eA}$ of an action $A$ is defined as follows:

$$t_{eA} = \max \{ p_T(t) ; e \in A, \text{ and } t \text{ is the tag of } e\}$$

**Definition 2.37.** a) Two actions are called *separable* when the end of one of them is less than or equal to the beginning of the other.

b) Two actions are called *strictly separable* when the end of one of them is less than the beginning of the other.

**Definition 2.38.** A process $P$ is called *action sequential* if all actions from $A$ are separable from one another in the sense of

**Definition 2.39.** A process $P$ is called *strictly sequential* if all actions from $P$ are sequential and separable from one another.

**Proposition 2.12.** Any subprocess of an action sequential (strictly sequential) process of $P$ is an action sequential (strictly sequential) process.

**Definition 2.40.** A process $P$ is called *interleaving* if all events from $P$ have different time, i.e., if $e_i = (d_i, t_i)$ and $e_j = (d_j, t_j)$ are events from $A$, then $p_T(t_i) = p_T(t_j)$ implies $e_i = e_j$.

**Proposition 2.13.** Any subprocess of an interleaving process of $P$ is an interleaving process.

**Proposition 2.14.** Any interleaving process of $P$ that has only point-like events is strictly sequential.

For processes with extended and/or pointed events, this result in general is not true.

## 3. Algebras of Events, Actions, and Processes

### 3.1. Algebraic Foundations for Modeling Concurrency

As in our model, we have three levels of description of concurrent processes, formalization of this model brings us to three-level algebras.

**Definition 3.1.** A three-level algebra $\mathbf{M} = (\mathbf{A}, \mathbf{B}, \mathbf{C}, \mathbf{H}, \mathbf{K})$ consists of three horizontal algebras $\mathbf{A}$, $\mathbf{B}$, and $\mathbf{C}$ called *levels* and two vertical algebras $\mathbf{H}$ and $\mathbf{K}$ called *lifts*:

1. The first level is $\mathbf{A} = (G_A, Op_A, A)$ where $G_A$ is a set of generators of the algebra $\mathbf{A}$, $Op_A$ is the set of all operations of the algebra $\mathbf{A}$, and $A$ is the support (set of all elements) of the algebra $\mathbf{A}$. Operations from $Op_A$ are called inner or internal operations of the algebra $\mathbf{A}$, or horizontal operations of the algebra $\mathbf{M}$.

2. The first lifting algebra is $\mathbf{H} = (A, O^p, B)$ where $A$ is the support of the algebra $\mathbf{A}$, $O^p$ is the set of all operations of the algebra $\mathbf{H}$, and $B$ is the support of the algebra $\mathbf{B}$. Operations from $O^p$ are called inner or internal operations of the algebra $\mathbf{H}$, or vertical operations of the algebra $\mathbf{M}$. Operations from $O^p$ are applied to elements from $A$ to build elements from $B$ and we assume that all elements from $B$ are built in such a way.

Vertical operations are graded operations similar to operations in graded algebras (Higgins, 1963).

3. The second level is $\mathbf{B} = (G_B, Op_B, B)$ where $G_B$ is a set of generators of the algebra $\mathbf{B}$, $Op_B$ is the set of all operations of the algebra $\mathbf{B}$, and $B$ is the support of the algebra $\mathbf{B}$. Operations from $Op_A$ are called inner or internal operations of the algebra $\mathbf{B}$, or horizontal operations of the algebra $\mathbf{M}$.

4. The second lifting algebra is $\mathbf{K} = (B, O^{pp}, C)$ where $B$ is the support of the algebra $\mathbf{B}$, $O^{pp}$ is the set of all operations of the algebra $\mathbf{K}$, and $C$ is the support of the algebra $\mathbf{C}$. Operations from $O^p$ are called inner or internal operations of the algebra $\mathbf{H}$, or vertical operations of the algebra **M**. Operations from $O^{pp}$ are applied to elements from $B$ to build elements from $C$ and we assume that all elements from $C$ are built in such a way.

5. The third level is $\mathbf{C} = (G_C, Op_C, C)$ where $G_C$ is a set of generators of the algebra **C**, $Op_C$ is the set of all operations of the algebra **C**, and $C$ is the support of the algebra **C**. Operations from $Op_A$ are called inner or internal operations of the algebra **C**, or horizontal operations of the algebra **M**.

We assume that all operations and consequently, some of these algebras can be partial. However, building algebras of events, actions, and processes, we do not specify when operations and corresponding algebras are total or partial. For instance, a monoid or a group is a total algebra, while a category is a partial algebra. Often we consider total operations (total algebras) as particular cases of partial operations (partial algebras).

Note that three-level algebras, like ordinary (universal) algebras, can have additional structures, such as topology or order (Bourbaki, 1960; Kurosh, 1963).

Three levels and two lifts give us the following partially commutative diagram:

$$
\begin{array}{ccc}
G_A & \xrightarrow{Op_A} & A \\
{\scriptstyle O^p}\downarrow & & \downarrow{\scriptstyle O^p} \\
G_B & \xrightarrow{Op_B} & B \\
{\scriptstyle O^{pp}}\downarrow & & \downarrow{\scriptstyle O^{pp}} \\
G_C & \xrightarrow{Op_C} & C
\end{array}
\qquad (1)
$$

Taking sets $Op_B^{ind}$ and $Op_C^{ind}$ of operations induced by operations from $Op_A$ and $Op_B$ in algebras **B** and **C** instead of sets $Op_B$ and $Op_C$, correspondingly, in the diagram (1), we obtain two commutative squares (2) and (3).

$$\begin{array}{ccc} G_A & \xrightarrow{Op_A} & A \\ O^p \downarrow & & \downarrow O^p \\ G_B & \xrightarrow{Op_B^{ind}} & B \end{array} \quad (2)$$

$$\begin{array}{ccc} G_B & \longrightarrow & B \\ O^{pp} \downarrow & & \downarrow O^{pp} \\ G_C & \xrightarrow{Op_C^{ind}} & C \end{array} \quad (3)$$

In a natural way, homomorphisms of three-level algebras are defined.

Three-level algebras are generalizations of graded algebras in the sense of (Higgins, 1963).

## 3.2. General principles for compositions of processes and actions

Usually only two compositions of processes, also called operations with processes, are considered: parallel and sequential composition. In practice, for example, in concurrent programming languages more compositions have been developed. In

(Burgin and Smith, 2006, 2006a), it is demonstrated that there are even several kinds of sequential composition.

As our goal is to build algebras of processes, we need to determine enough compositions of processes to be able to model real situations that exist in concurrent computations, communications and networking.

We start with developing a typology of process compositions. This typology will give us a plan for designing flexible and efficient constructions of process compositions. Here we consider only binary compositions. Ternary and higher arity compositions are considered elsewhere.

Temporal types of processes and actions compositions:
1. Sequential compositions.
2. Concurrent compositions.
3. Multiscale compositions.

Sequential compositions of processes have three subtypes:
1. Consequential compositions.
2. Action interleaving compositions.
3. Event interleaving compositions.

Concurrent compositions of processes and actions have three subtypes:
1. Free compositions.
2. Parallel compositions.
3. Partially parallel compositions.

Information types of processes and actions compositions:
1. Compositions without interactions.
2. Interactive compositions.
3. Transformative compositions.

Interactive compositions are built to reflect interactions between processes and actions without taking into account substantial and informational content of interactions. Transformative compositions are built to organize interactions between processes and actions and represent not only how this interaction goes but also what is exchanged between processes and actions. In particular, information transmission and transformation is explicitly described.

According to the form, there are two types of composition:

1. Compositions with explicit interactions.
2. Compositions with implicit interactions.
3. Compositions with both types of interaction.

It is necessary to remark that algebras of processes have unary operations and operations with arity more than 2, i.e., applied to more than two elements (events, actions or processes), in addition to various types of binary compositions. Some operations are total, while others are partial, i.e., undefined for some elements of these algebras. Some operations that have arity more than 2 are generated by binary compositions, while others cannot be reduced to binary compositions.

### 3.3. Compositions of Events

As systems $V$ and $ST$ have topological and algebraic structures, these structures induce corresponding structures in sets $\mathbf{EV}_{NV, ST}$, $\mathbf{EV}_{NV, EST}$, and $\mathbf{EV}_{NV, kST}$ of events.

For instance, the semantic system $V$ is usually an algebraic system. This means that there are (in general, partial) operations in $V$. These operations induce corresponding operations in the system of names $N$.

This allows us to define such a partial binary operation in the space $\mathbf{EV}_{NV,ST}$ of point-like events as the direct parallel composition of point-like events.

Let $m$ be a binary operation in $N$, and $e_1 = (d_1, t)$ and $e_2 = (d_2, t)$ be parallel point-like events. When the semantic function *sem* is one-to-one, operations in $N$ are usually induced by operations in $V$.

**Definition 3.2.** The (*direct parallel*) $m$-composition $m(e_1, e_2)$ of events $e_1$ and $e_2$ is defined by the formula

$$m(e_1, e_2) = (m(d_1, d_2), t)$$

**Example 3.1.** Parallel composition of information processing events is used for constructing an operation of a Turing machine from such events as changing the state of the machine, writing a symbol in a cell, erasing a symbol from a cell, and moving the head of the machine.

**Example 3.2.** In MIMD computing architecture, several functional units perform different operations on different data. Examples are a multiprocessor or transputer computer, or a network of workstations. In this case, an event is a combined operation of several functional units. This event is a composition of operations in each separate unit. Similar operations on events are organized in SIMD computing architecture.

When both systems $N$ and $ST$ are metric spaces with distances $\mathbf{d}_N$ and $\mathbf{d}_{ST}$, correspondingly, it is possible to introduce a metric in $\mathbf{EV}_{NV,ST}$, making $\mathbf{EV}_{NV,ST}$ a metric space.

One way to define a metric in $\mathbf{EV}_{NV,ST}$, is to use the Euclidean extension of metrics. Namely, if we take two events $e_1 = (d_1, t_1)$ and $e_2 = (d_2, t_2)$ from $\mathbf{EV}_{NV,ST}$, then the distance between them is equal to $\mathbf{d}(e_1, e_2) = \sqrt{\mathbf{d}_N^2(d_1, d_2) + \mathbf{d}_{ST}^2(t_1, t_2)}$.

Another way to do this, is to use the Shannon extension of metrics. Namely, if we take two events $e_1 = (d_1, t_1)$ and $e_2 = (d_2, t_2)$ from $\mathbf{EV}_{NV,ST}$, then $\mathbf{d}(e_1, e_2) = \mathbf{d}_N(d_1, d_2) + \mathbf{d}_{ST}(t_1, t_2)$.

To develop compositions of events, we assume that the semantic system $V$ is an algebraic system. This means that there are (in general, partial) operations in $V$.

**Proposition 3.1.** a) $m(e_1, e_2) \parallel e_1 \parallel e_2$.  b) If $e_1 \parallel e_2 \parallel e_3 \parallel e_4$, then $m(e_1, e_2) \parallel m(e_3, e_4)$ and $m(e_1, e_2) \parallel e_3 \parallel e_4$.

A useful operation with events is a shift.

Let *sft*: $ST \rightarrow ST$ be an arbitrary mapping and $e = (v, t)$ be an event from $\mathbf{EV}_{V,ST}$.

**Definition 3.3.** The *sft*-shift $F_{sft}(e)$ of the events is equal to $(v, sft(t))$.

**Proposition 3.2.** If two events are parallel, then the results of the *sft*-shift of these events are parallel.

Let *sft*: $ST \to ST$ be a mapping such that $sft(t_1) = t_2$ and $e_1 = (v_1, t_1)$ and $e_2 = (v_2, t_2)$ be events from $\mathbf{EV}_{V, ST}$.

**Definition 3.4.** The shifted *m*-composition $m_{sft}(e_1, e_2)$ of events $e_1$ and $e_2$ is defined by the formula

$$m_{sft}(e_1, e_2) = (m(d_1, d_2), t_2)$$

Informally, shifted *m*-composition means that at first we shift the first event to the same position in time and space that has the second event and then apply the direct parallel composition.

Definition 3.3 implies that $m_{sft}(e_1, e_2)$ is always parallel to $e_2$.

**Proposition 3.3.** If $e_2 \parallel e_4$, then $m_{sft}(e_1, e_2) \parallel m_{sft}(e_3, e_4)$.

Indeed, validity of the relation of parallelism $\parallel$ depends only on the time projection of the event tags, while the shifted *m*-composition changes only names of events and takes the tag of the second event from this composition. It means that time projections of both compositions coincide with projections of the corresponding second arguments in these compositions and by the initial condition this time is the same for both second arguments as they are parallel.

A semantic shift of events is called *renaming* as it is possible to consider values of events as names of these events.

Let *rn*: $V \to V$ be an arbitrary mapping and $e = (v, t)$ be an event from $\mathbf{EV}_{V, ST}$.

**Definition 3.5.** The *rn*-renaming $R_{rn}(e)$ of the events is equal to $(rn(v), t)$.

**Proposition 3.4.** If $e_1 \parallel e_2$, then $R_{rn}(e_1) \parallel R_{rn}(e_2)$.

Indeed, validity of the relation of parallelism $\parallel$ depends only on the time projection of the event tags, while renaming changes only names of events.

### 3.4. Compositions of Actions

There is a diversity of compositions of actions and even larger diversity of compositions of processes. Many of these compositions are used in practice. Our goal here is to develop a theoretical foundation for communication and computation in technical and other systems, as well as to study these compositions with the aim to provide new efficient means for communication and computing practice.

**Definition 3.6.** The *free composition* of actions $A_1$ and $A_2$ is the action $A$ that consists of all events from $A_1$ and $A_2$ taken without changes and without introduction of new relations.

The free composition of actions $A_1$ and $A_2$ is denoted by $A_1 \cup A_2$.

Note that the free composition of actions $A_1$ and $A_2$ is not simply the union of two sets $A_1$ and $A_2$ because the free composition takes into account existing relations between events in $A_1$ and $A_2$.

**Definition 3.7.** The *meet* of actions $A_1$ and $A_2$ is defined as the action that consists of all events that are elements of both $A_1$ and $A_2$ and all relations between these events that belong both to $A_1$ and $A_2$.

The meet of actions $A_1$ and $A_2$ is denoted by $A_1 \cap A_2$.

Note that the meet of actions $A_1$ and $A_2$ is not simply the intersection of two sets $A_1$ and $A_2$ because the meet takes into account existing relations between events in $A_1$ and $A_2$.

Let **E** be some set of events and A(**E**) be the set of all actions generated by the events from **E**, i.e., elements of action from A(**E**) are events from **E** and only such events.

**Theorem 3.1.** a) The free composition of actions is an idempotent, commutative and associative operation, i.e., the set A(**E**) is an Abelian (commutative) monoid with the empty action as its identity element, as well as a semilattice.

b) The set A(**E**) is a Boolean algebra with respect to operations $\cup$ and $\cap$.

To build other compositions of actions, we use shifts of actions. Any *shift of an action* is constituted of the shifts of events comprised in this action.

**Proposition 3.5.** The result of any shift of an action $A$ without repetition is an action without repetition.

For coordinated and sequential actions, this is not true in general. However, in some cases, we have the same result.

**Definition 3.8.** A shift of an action $A$ is *uniform* if the shifts of all events from $A$ are defined by the same mapping $sft: ST \to ST$.

**Proposition 3.6.** The result of a uniform shift of a coordinated (sequential) action $A$ is a coordinated (correspondingly, sequential) action.

**Proposition 3.7.** The result of any renaming of a sequential action $A$ is a sequential action.

For coordinated actions and actions without repetition, this is not true in general. However, in some cases, we have the same result.

**Proposition 3.8.** If $rn: V \to V$ is an injection, then the result of the corresponding renaming of a coordinated action (action without repetition) is a coordinated action (correspondingly, action without repetition).

**Definition 3.9.** A *free sequential composition* of actions $A_1$ and $A_2$ is obtained by the following operations:

- at first, we shift, if necessary, all events in $A_2$ with a shift $sft$, so that for any events $e_i$ from $A_1$ and $e_j$ from $A_2$, time in the shifted event $F_{sft}(e_j)$ is larger than time in the shifted event $e_i$;
- then we take the free composition of these actions $A_1$ and $F_{sft}(A_2)$.

The free sequential composition of actions $A_1$ and $A_2$ is denoted by $A_1 \cdot A_2$.

If the first process is finite, then its free sequential composition with any other process is defined. Note that the case $i = j$ is also included in the definition.

**Proposition 3.9.** The free sequential composition of two sequential actions is a sequential action.

**Definition 3.10.** A *free interleaving composition* of two actions $A_1$ and $A_2$ is obtained by the following operations:

- at first, we shift, if necessary, all events in $A_2$ with a shift *sft*, so that for any events $e_1$ from $A_1$ and $e_2$ from $A_2$, time in the shifted event $F_{sft}(e_2)$ is not equal to time in the shifted event $e_1$;

- then we take the free composition of these actions $A_1$ and $F_{sft}(A_2)$.

The free interleaving composition of actions $A_1$ and $A_2$ is denoted by $A_1 \Theta A_2$.

**Proposition 3.10.** The free interleaving composition of two sequential actions is a sequential action.

Let $W \subseteq V$.

**Definition 3.11.** The $W$-projection $Pr_W(A)$ of an action $A$ is obtained by excluding from $A$ all events with values not in $W$.

Let $K \subseteq ST$.

**Definition 3.12.** The $K$-projection $Pr_K(A)$ of an action $A$ is by excluding from $A$ all events with tags not in $K$.

**Lemma 3.1.** Any projection of an action $A$ is a subaction of $A$.

**Theorem 3.2.** Any projection of a coordinated action (action without repetition or sequential action) $A$ is a coordinated action (respectively, action without repetition or sequential action).

**Definition 3.13.** A *free parallel composition* of two actions $A_1$ and $A_2$ is obtained by the following operations:

- at first, we shift some events in $A_2$ with a shift *sft*, so that all parallel events in $A_2$ preserve their parallelism and some events in $F_{sft}(A_2)$ become parallel to some events in $A_1$;

- then we take the free composition of these actions $A_1$ and $F_{sft}(A_2)$.

The free parallel composition of actions $A_1$ and $A_2$ is denoted by $\pi(A_1, A_2)$.

We remind the reader that as we consider only point-like events, their parallelism means that they happen simultaneously. Usually parallel composition has definite restrictions. These restrictions can come from:

- realization of actions and events, e.g., if we have ten processors, we cannot simultaneously perform eleven operations when each operation demands one processor;
- the algorithm, e.g., we cannot simultaneously perform two operations if one of them uses the result of the other one.

**Proposition 3.11.** If $\pi$ is a free parallel composition of actions $A$ and $B$, then

$$m^{par}(\pi(A, B)) > \max \{ m^{par}(A), m^{par}(B) \}$$

and

$$m_{par}(\pi(A, B)) \geq m_{par}(A) + m_{par}(B).$$

### 3.5. Compositions of Processes

The most basic composition of processes is their free composition.

**Definition 3.14.** The *free composition* of processes $P_1$ and $P_2$ is the process $P$ that consists of all actions from $P_1$ and $P_2$ taken without changes and introduction of new relations.

The free composition of actions processes $P_1$ and $P_2$ is denoted by $P_1 \cup P_2$.

Note that the free composition of processes $P_1$ and $P_2$ is not simply the union of two sets $P_1$ and $P_2$ because the free composition takes into account existing relations between actions in $P_1$ and $P_2$.

**Definition 3.15.** The *meet* of processes $P_1$ and $P_2$ is defined as the process $Q$ that consists of all actions that are elements of both $P_1$ and $P_2$ and all relations between these actions that belong both to $P_1$ and $P_2$.

The meet of processes $P_1$ and $P_2$ is denoted by $P_1 \cap P_2$.

Note that the meet of $P_1$ and $P_2$ is not simply the intersection of two sets $P_1$ and $P_2$ because the meet takes into account existing relations between events in $P_1$ and $P_2$.

Let **A** be some set of actions and P(**A**) be the set of all actions generated by the actions from **A**, i.e., elements of action from P(**A**) are events from **A** and only such events.

**Theorem 3.3.** a) The free composition of processes is an idempotent, commutative and associative operation, i.e., the set P(**A**) is an Abelian (commutative) monoid with the empty process as its identity element, as well as a semilattice.

b) The set P(**A**) is a Boolean algebra with respect to operations ∪ and ∩.

To build other compositions of processes, we use shifts of processes. Any *shift of a process* is constituted of the shifts of events comprised in this process.

**Proposition 3.12.** The result of any shift of a process $P$ without repetition is a process without repetition.

For coordinated and sequential processes, a similar statement is not true in general. However, in some cases, we have the same result.

**Definition 3.16.** A shift of a process $P$ is *uniform* if the shifts of all events from $P$ are defined by the same mapping *sft*: $ST \to ST$.

**Proposition 3.13.** The result of a uniform shift of a coordinated (sequential) process $P$ is a coordinated (correspondingly, sequential) process.

**Proposition 3.14.** The result of any renaming of a sequential process $P$ is a sequential process.

For coordinated processes and processes without repetition, this is not true in general. However, in some cases, we have the same result.

**Proposition 3.15.** If *rn*: $V \to V$ is an injection, then the result of the corresponding renaming of a coordinated process (process without repetition) is a coordinated process (correspondingly, process without repetition).

**Definition 3.17.** A *free sequential composition* of processes $P_1$ and $P_2$ is obtained the following operations:

- at first, we shift, if necessary, all events in $P_2$ with a shift *sft*, so that for any events $e_i$ from $P_1$ and $e_j$ from $P_2$, time in the shifted event $F_{sft}(e_j)$ is larger than time in the shifted event $e_i$ ;

- then we take the free composition of these processes $P_1$ and $F_{sft}(P_2)$.

The free sequential composition of processes $P_1$ and $P_2$ is denoted by $P_1 \cdot P_2$.

If the first process is finite, then its free sequential composition with any other process is defined. Note that the case $i = j$ is also included in the definition.

**Proposition 3.16.** The free sequential composition of two sequential processes is a sequential process.

**Definition 3.18.** A *free interleaving composition* of two processes $P_1$ and $P_2$ is obtained by the following operations:

- at first, we uniformly shift, if necessary, all actions in $P_2$ with a shift *sft*, so that for any events $e_1$ from an action $A_1$ from $P_1$ and $e_2$ from an action $A_2$ from $P_2$, time in the shifted event $F_{sft}(e_2)$ is not equal to time in the shifted event $e_1$ ;

- then we take the free composition of these processes $P_1$ and $F_{sft}(P_2)$.

The free interleaving composition of processes $P_1$ and $P_2$ is denoted by $P_1 \Theta P_2$.

**Proposition 3.17.** The free interleaving composition of two sequential processes is a sequential process.

Let $W \subseteq V$.

**Definition 3.19.** The *W-projection* of a process $P$ is obtained by excluding from $A$ all events with values not in $W$.

Let $K \subseteq ST$.

**Definition 3.20.** The *K-projection* of a process $P$ is by excluding from $P$ all events with tags not in $K$.

**Lemma 3.2.** Any projection of a process $P$ is a subprocess of $P$.

**Theorem 3.4.** Any projection of a coordinated process (process without repetition or sequential process) $P$ is a coordinated process (respectively, process without repetition or sequential process).

**Definition 3.21.** A *free parallel composition* of two processes $P_1$ and $P_2$ is obtained by:

- at first, shifting some events in $P_2$ with a shift *sft*, so that all parallel events in $P_2$ preserve their parallelism and some events in $F_{sft}(P_2)$ become parallel to some events in $P_1$;
- then taking the free composition of these processes $P_1$ and $F_{sft}(P_2)$.

We remind the reader that as we consider only point-like events, their parallelism means that they happen simultaneously. Usually parallel composition has definite restrictions. These restrictions can come from:

- realization of processes and events, e.g., if we have ten processors, we cannot simultaneously perform eleven operations when each operation demands one processor;
- from the algorithm, e.g., we cannot simultaneously perform two operations if one of them uses the result of the other one.

**Proposition 3.18.** If $\pi$ is a parallel composition of processs $P$ and $B$, then

$$m^{par}(\pi(P, Q)) > \max \{ m^{par}(P), m^{par}(Q) \}$$

and

$$m_{par}(\pi(P, Q)) \geq m_{par}(P) + m_{par}(Q).$$

### 3.6. Algebras of Processes

For processes, we build the following three-level process algebra **Pr** = (**Ev**, **Ac**, **Pr**, *EA*, *AP*) in EAP.

1. The first level **Ev** = (*EE*, *Op$_E$*, *Ev*) is an algebra of events. In it, *EE* is a set of elementary events in some system *R*, which are generators of the algebra **Ev**, *Op$_E$* is the set of all operations of the algebra **Ev**, and *Ev* is a set of events in the system *R*, which form the support (set of all elements) of the algebra **Ev**. Operations from *Op$_E$* are called inner or internal operations of the algebra of events **Ev**, or horizontal operations of the process algebra **Pr**.

The system of operations in the algebra **Ev** contains oprations on events described in the previous section: the direct parallel *m*-composition $m(e_1, e_2)$, strong sequential composition $e_1 \circ e_2$, and shifted *m*-composition $m_{sft}(e_1, e_2)$ of events, *sft*-shift $F_{sft}(e)$ and *rn*-renaming $R_{rn}(e)$ of an event.

2. The first lifting algebra is **EA** = (*Ev*, $O^p$, *Ac*) where *A* is the support of the algebra **A**, $O^p$ is the set of all operations of the algebra **H**, and *B* is the support of the algebra **B**. Operations from $O^p$ are called inner or internal operations of the algebra **H**, or vertical operations of the process algebra **Pr**. Operations from $O^p$ are applied to elements from *Ev* to build elements from *Ac* and we assume that all elements from *B* are built in such a way.

3. The second level **Ac** = (*Ac*, $Op_{Ac}$, *Ac*) is an algebra of actions. In it, $Op_A$ is the set of all operations in actions and *Ac* is a set of actions in the system *R*, which form the support of the algebra of actions **Ac**. Operations from $Op_{Ac}$ are called inner or internal operations of the algebra of actions **Ac**, or horizontal operations of the process algebra **Pr**.

The system of operations in the algebra **Ac** contains oprations on actions described in the previous section: the free composition $A_1 \cup A_2$, meet $A_1 \cap A_2$, free sequential composition $A_1 \cdot A_2$, free interleaving composition $A_1 \Theta A_2$, strong sequential composition $A_1 \circ A_2$, free parallel composition $\pi(A_1, A_2)$ of actions, and *W*-projection $Pr_W(A)$, *K*-projection $Pr_K(A)$, *sft*-shift $F_{sft}(A)$ and *rn*-renaming $R_{rn}(A)$ of an action.

4. The second lifting algebra is **K** = (*B*, $O^{pp}$, *Pr*) where *B* is the support of the algebra **B**, $O^{pp}$ is the set of all operations of the algebra **K**, and *C* is the support of the algebra **C**. Operations from $O^p$ are called inner or internal operations of the algebra **H**, or vertical operations of the process algebra **Pr**. Operations from $O^{pp}$ are applied to elements from *B* to build elements from *C* and we assume that all elements from *C* are built in such a way.

5. The third level **Pr** = (*Pr*, $Op_{Pr}$, *Pr*) is an algebra of processes. In it, $Op_{Pr}$ is the set of all operations in processes and *Pr* is a set of processes in the system *R*, which form the support (set of all elements) of the algebra of processes **Pr**. Operations from $Op_{Pr}$ are called inner or internal operations of the algebra of processes **Pr**, or horizontal operations of the process algebra **Pr**.

The system of operations in the algebra **Pr** contains oprations on processes described in the previous section: the free composition $P_1 \cup P_2$, meet $P_1 \cap P_2$, free sequential composition $P_1 \cdot P_2$, free interleaving composition $A_1 \Theta A_2$, free parallel composition $\pi(A_1, A_2)$ of processes, and *W*-projection $Pr_W(P)$, *K*-projection $Pr_K(P)$, *sft*-shift $F_{sft}(P)$ and *rn*-renaming $R_{rn}(P)$ of a process.

Some operations in events, actions, and processes are total, while others are partial.

## 4. EAP and other Concurrency Models

**4.1. Comparing EAP with ESP**

Lee and Sangiovanni-Vincentelli (1996) describe a concurrency metamodel, ESP, the basic elements of which are events, signals, and processes. ESP is intended to provide a framework for comparing other existing computational models in terms of how concurrency, communication, and time are represented. In this subsection, we demonstrate that the ESP metamodel is a special case of our more general EAP metamodel. Specifically, we describe how it is possible within EAP to reason about all the models of concurrency one can use ESP to reason about. In the process, it will become evident that EAP may be also used to reason about concurrent systems that the

ESP metamodel does not readily support. This shows greater descriptive and cognitive power of EAP in comparison with ESP.

To represent basic constructs from ESP in EAP, we start with events. Namely, the set of all events in ESP is corresponded to a subset $EV_{ESP}$ of the set $\mathbf{EV}_{NV,\,ST}$ of point-like events, which we defined in Section 2. ESP events are tag-value pairs where the value belongs to a set V and represents an observable event, while the tag belongs to a set T. In other words, $T \times V$, is the set of all events from ESP. The value of an event describes what happens in this event, while the tag indicates the instant in time when the corresponding event occurred. Thus, we need to take a subset $EV_{ESP}$ such that $N = V = V$ and $ST = T$, while the semantic function *sem* and the natural projection $p_T$ are identity mappings. These conditions make the set $T \times V$ of all events in ESP and the set $EV_{ESP}$ isomorphic and provide a representation of all events from ESP in the set $EV_{ESP}$ from EAP.

It is necessary to remark that, as Lee and Sangiovanni-Vincentelli write (1996), tags of events from ESP, in a general case, are able to indicate not only time. However, this is also true for tags in EAP, which contain the space coordinate in addition to the time coordinate and can belong to a broader structure than the space-time manifold.

Events in ESP are used to build signals, the basic construct from the next level of ESP. These signals are subsets of the direct product $T \times V$ and may be represented as sets or tuples of events. The set S of all signals from ESP naturally coincides with the powerset of $T \times V$. If *n* is a natural number, then the set containing all collections (or tuples) of signals of size *n* is denoted $S^n$. At the same time, the aggregations of events in EAP are actions, and the signals of ESP are one type of actions where all events are from the set $EV_{ESP}$ and there are no additional relations between these events in the corresponding action. As a result, these conditions and the restriction of ESP's point-like events considered above reduces EAP's actions to signals within ESP.

Signals in ESP are used to build processes, the basic construct from the top level of ESP. A process, in ESP, is an *n*-tuple of signals, that is, an element from $S^n$ for some *n*.

Signals in these processes are used for interactions between processes as Lee and Sangiovanni-Vincentelli remark (1996). Thus, any process from ESP can be represented by such a process in EAP in which all actions are signals and include only point-like events. This shows that the EAP metamodel completely encompasses the ESP metamodel.

However, EAP provides much more possibilities for modeling real concurrent processes in various systems: in computers and computer networks, in embedded systems and biological organism, and so on. For example, the inclusion of events with duration permits actions that cannot be represented by ESP's signals. For example, ESP's signals may include only events that occur either sequentially or simultaneously, but EAP's actions may also include partially overlapping events. One more important feature of big dynamical systems (either physical systems or global networks) is time relativity. In the ESP metamodel, relativistic effects are neglected and it is assumed that time is the same everywhere. In contrast to this, the EAP metamodel base its constructions on the space-time manifold *ST*, which allows one to take into account relativistic and other effects that are inherent for big dynamical systems.

Beyond the differences in temporal relationships between events that distinguish actions from signals, there is another significant difference in meaning between these two middle-tier components of ESP and EAP. Signals from ESP represent possible behaviors of a process. That is, a signal represents a possible trace of a process, and by extension, a process under ESP is defined by its set of possible behaviors, or simply, a set of signals. In contrast, the actions of EAP do not always represent behaviors of one process, per se, but rather permit reasoning about crosscutting multi-process and sub-process behavior, e.g., threads and fibers. Thus, EAP actions permit reasoning about process interactions directly, and are more than an intermediate level of composition for processes. This crosscutting middle tier is an important contribution of EAP. In addition, three types of actions, actualized, potential and emerging, allow one not only to represent possible behaviors of a system, but also to discern what really happened from what can tentatively be.

Demonstrating that EAP models all possibilities of ESP to reason about concurrent processes and taking into account transitivity of modeling relation, we have as a consequence that EAP can represent and encompass the cluster of concurrency models that are comprised by ESP. As it is establishted in (Lee and Sangiovanni-Vincentelli, 1996), this cluster contains such models as Kahn process networks (Kahn, 1974), the CSP model of Hoare (1985), the CCS model of Milner (1989), the dataflow model (Lee and Parks, 1995), Petri nets (Petri, 1962; Peterson, 1991; Reisig, 1985) and other discrete-event simulators.

ESP's definition of process is consistent with the CSP Traces model, and the basis for the CSP's process calculus (Hoare, 1985). That is, one can reason about properties of a process by reasoning about all possible behaviors of a process. Moreover, ESP's definition of events support directly reasoning about parallel events in the spirit of View-Centric Reasoning (VCR) by Smith (2000), which itself was an extension of CSP. However, VCR goes further to provide the possibility of multiple, possibly imperfect observers, and by extension, multiple corresponding views of the same computation. These views permit reasoning about ambiguities such as occur in the Linda model when Linda predicate operations rdp() and inp() indicate a failure to match a tuple in Tuple Space. This scenario is discussed in greater detail in Smith (2000). Without multiple perspectives (views) of a computation's history, it is difficult to appreciate the subtlety and subsequent disambiguation of these operations in the case of failure.

### 4.2. Comparing EAP with VCR

In this section we compare EAP with View-Centric Reasoning (VCR). VCR arose from the operational semantics for concurrency developed by Smith (2000). This operational semantics, in addition to providing a description of how computation proceeds for generic concurrent systems, also afforded means for constructing parallel event traces similar to Hoare's construction of event traces in CSP (Hoare 1985). Thus,

VCR traces are derived from an operational semantics for concurrency, while CSP traces are derived from a process algebra. We now relate the process algebra of CSP and operational semantics of VCR, and along the way, provide means for comparison between VCR and EAP.

In Smith (2000), a parameterized, parallel and distributed operational semantics (paraDOS) is developed as a meta model for reasoning about concurrency. Moreover, paraDOS form the basis for View-Centric Reasoning (VCR). The first models of concurrency paraDOS has been instantiated for are Actors and Linda. The Actors model and the Linda model were chosen to be instantiated because they represent very different models of concurrency, and modeling both was able to demonstrate the genericity of paraDOS and VCR.

Before describing the components of paraDOS in more detail, it is helpful to describe the two models that were instantiated: Actors and Linda. This will permit a better understanding of paraDOS, which provide the means of this instantiation. The Actors model is due to Agha (1986), with later development by Agha, et al (1997) and Mason and Talcott (1997). The Linda model is due to Gelernter (1985), with later pertinent development by Carriero and Gelernter (1993). Both models are poised to play important roles in harnessing the computational power of multicore processor architectures, which provides an added dimension of relevance to being able to reason about concurrent systems based on these models.

Let us briefly consider the Actors model. In it, an actor is represented by its mail queue, and an actor machine is an instance of an actor's behavior (i.e., of a continuation). An actor machine is bound to one (and only one) element of the actor's mail queue. An actor machine's behavior is a function of the message it receives from its mail queue. Actors communicate by sending messages (asynchronously) to another actor's mail queue. Message delivery is guaranteed (in the limit), though order of delivery is not. Thus, the Actors model is super-recursive (cf. (Burgin, 2005)) because message delivery is guaranteed only in the limit. As a result, the time and order of message delivery are the major sources of nondeterminism in the Actors model of concurrency.

The Linda model, in contrast, is not based on message passing, but neither it is a classic shared memory model. Gelernter (1985) classified the Linda model as a new communication paradigm, describing it as generative communication. Linda is noted for its communication orthogonality, with communications being decoupled in three dimensions: destination (anonymous "receivers"), space (heterogeneous communication), and time (time-disjoint interprocess communication). Linda processes share access to a distributed, associative memory called tuple space, where tuples may be placed, copied, and removed by the Linda language primitives out(), rd(), and in(), respectively. Linda is a coordination and coordination language that augments an existing computational language. Sources of nondeterminism in Linda arise when a single Linda process could match multiple tuples, or when multiple simultaneous processes vie for the same tuple.

In designing paraDOS as a metamodel for concurrency, Smith (2000) observed that all concurrent systems could be characterized as consisting of processes and interprocess communications. The operational semantics of paraDOS required identifying the components of a system's state, and defining a transition relation between states. What resulted was a state that consists of a 4-tuple containing a set of process behaviors, a set of communication closures, a parallel event set, and, recursively, the next state to which computation proceeds. The set of process behaviors is represented by continuations. The communication closures are bound expressions, which serve as a unifying abstraction for a variety of communication paradigms, such as shared memory and message passing. The parallel event set contains the multiset of observable events that occurred in transitioning from the previous state to the current state. The parallel event set permits constructing traces of a system's computation, and it is this trace that connects paraDOS and VCR to both CSP and EAP. The final element of the 4-tuple is the recursive next state, which is lazy in that it does not exist until the transition relation elaborates it (i.e., the transition relation chooses which state to elaborate from among the set of all possible next states).

Thus, to instantiate paraDOS for Actors and Linda, one must define the components that collectively make up the state, and the transition relation between states. An important part of this instantiation process is identifying which events are observable as a system transitions from state to state. For the Actors instantiation of paraDOS, observable events are the sending and delivery of messages to an actor's mail queue. For the Linda instantiation of paraDOS, observable events include the creation, copying, and consuming of tuples in tuple space. More details on the instantiation of paraDOS for Actors and Linda are presented in (Smith, 2000). In both instances of paraDOS, the observable events represent stages of communication progress.

In two theorems proved in (Smith, 2000) equivalence between Actors, Linda and corresponding restricted versions of paraDOS is established. Namely, the first theorem demonstrates equivalence of a restricted version of paraDOS for Actors to the *Actor Theory* operational semantics introduced by Mason and Talcott (Mason & Talcott, 1997). The second theorem demonstrates equivalence of a restricted version of paraDOS for Linda to the *TSspec* operational semantics developed by Jensen (1994) and Ciancarini, et al. (1994). The restriction on paraDOS for both instantiations concerns one of its parameters, transition density, which was set to 1 to facilitate comparison with the Actor Theory and TSspec operational semantics, which were each limited to transitions involving only a single observable event. Without this restriction, as a model of true concurrency, paraDOS is capable of transitions involving multiple processes and multiple observable events. Exact statements of these theorems and respective proofs are given in (Smith, 2000).

The remainder of Section 4.2 describes how EAP can be used to model and reason about concurrent systems based on paraDOS, VCR, the Actors and Linda models.

Events from VCR can be exactly represented by point-like events in EAP. VCR events are instantaneous, and do not include tags, but this does not present a problem for modeling in EAP. One more feature of the EAP events that model VCR events is that tags of the EAP events represent only time and do not have space coordinates, i.e., they are pure temporal events (cf. Definition 2.6). In other words, equivalent EAP events occur in some sequence in time, but location information is not preserved. If this bothers the reader, one could consider events as all occurring in the same location.

VCR events constitute the observable evidence of state transitions in a particular instantiation of paraDOS. State transitions form computation. In this context, it is possible to consider VCR events as computational events. Likewise, in EAP, events describe transitions (Axiom E2a). Thus, each computational VCR event can be represented by an event in EAP by giving its description or value and its time coordinate. If necessary, it is also possible to use the space coordinate.

Processes in VCR/paraDOS are represented by their continuations. In general, it is possible to semantically represent a VCR/paraDOS process by an ordered set of computational events (not necessarily all observable). Taking actions that consist only of point-like events without space coordinate, we can represent all VCR processes by EAP processes that use only such actions. Such a representation of processes in EAP is equivalent to the use of continuations in paraDOS.

As a result, processes in VCR can be represented by processes in EAP, though we must restrict in EAP the type of events used to be point-like and without space coordinates. VCR's parallel event traces are naturally represented by EAP's actions, where parallel events from VCR are EAP point-like events with tags indicating that they occurred simultaneously.

The set of process continuations and the parallel event multiset from paraDOS induce in EAP a relation on events of being observable. Thus, being observable is a property of EAP events. We can map traces in VCR to actions in EAP by including in actions only those events that are observable. Since the act of observation of a computation affects the computation itself, it is appropriate to consider observation as an operation on events in the EAP process algebra. Such EAP actions, built from observable events, are *traceable actions*. In this way, VCR traces correspond precisely to EAP's traceable actions.

In such a way, we have demonstrated how the elements of EAP can be used to represent the elements of paraDOS and VCR. Since VCR has been shown to model such diverse concurrent models of concurrency as Actors and Linda, and EAP contains elements that model the element of VCR, we conclude that EAP is also capable of modeling the Actors and Linda models of concurrency.

### 4.3. Comparing operations in EAP with operations in other concurrency models.

All concurrency models have operators to combine processes and to form new processes from existing ones. Such operators provide representation of basic features of concurrent processes in process algebra: compositionality, concurrency, and communication. Let us consider the most important of these operators.

The most fundamental of algebraic operators are choice and sequential composition. The *alternative* operator + provides a choice between events/actions from the composed processes, and the *sequencing operator* · specifies the sequential ordering of events/actions in the two processes that are composed. For instance, the process $(P + Q) \cdot R$ first chooses to perform all events/actions either from the process $P$ or from the process $Q$, and then performs from the process $R$.

To allow the representation of concurrency, the *merge*, *right-merge*, and *left-merge* operators are used. The merge operator ∥ represents the parallel composition of two processes, the individual events/actions of which are interleaved. The left-merge operator ∥⌊ (also denoted by ⌊⌊) is an auxiliary operator with similar semantics to the merge, but a commitment to always choose its initial step from the left-hand process. The right-merge operator ⌋∥ (also denoted by ⌋⌋) is an auxiliary operator with similar semantics to the merge, but a commitment to always choose its initial step from the right-hand process. For instance, if we take two processes $U = (P \cdot Q)$ and $W = (R \cdot V)$ where $P$, $Q$, $R$ and $V$ are actions, then the process $U \parallel W = (P \cdot Q) \parallel (R \cdot V)$ may perform the actions *P, Q, R, V* in any of the sequences *PQRV, RPQV, RPVQ, RVPQ, PRQV, PRVQ*. At the same time, the process $(P \cdot Q) \parallel\!\!\!\!\!\bot\, (R \cdot V)$ may only perform the sequences *PQRV, PRQV, PRVQ* as the left-merge operator ensures that the actions *P* is performed first.

Interactions and, in particular, communications between processes are represented using the binary *communication operator* $|$. For example, let us consider the actions *r(d)* of reading a data item $d \in D$ and *w(d)* of writing a data item $d \in D$. Then the communication composition

$$(\sum\nolimits_{d \in D} r(d) \cdot y) | (w(1) \cdot z)$$

of processes $\sum\nolimits_{d \in D} r(d) \cdot y$ and $\sum\nolimits_{d \in D} w(d) \cdot y$ communicates the value 1 from the right component process to the left component process (i.e., the identifier *d* is assigned the value 1, and free instances of *d* in the process *y* take on that value), and then behave as the merge of *y* and z.

The abstraction operator $\tau_I$ provides a way to "hide" certain events, actions, and processes by treating them as unobservable events, actions, and processes going inside the system being modeled and thus, unrepresentable in the process algebra. Abstracted actions are converted to the *silent step* action $\tau$. In some cases, these silent steps can also be removed from the process expression. For instance, the abstraction $\tau_{\{c\}}((a + b) \cdot c) = (a + b) \cdot \tau$ allows one to reduce the result of abstraction to a + b since the event *c* is no longer observable and has no observable effects.

Compositions of processes in EAP determined by the compatibility relations allow us to represent all operators from different models of concurrency. Let us build such representations for the considered above basic operators of process algebras: sequencing, merge, right-merge, left-merge, communication and alternative operators.

The **alternative operator** *P + Q* of processes *P* and *Q* is obtained if we take the free composition of processes *P* and *Q* and add the compatibility relations that make possible either all actions from *P* or all actions from *Q*.

**Sequencing** of events, actions, and processes are represented by the free sequential composition of processes because sequencing of events or actions in models of concurrency gives a process as the result.

The **merge** $P \parallel Q$ of processes $P$ and $Q$ is obtained if we take the free composition of processes $P$ and $Q$ and add the compatibility relations that make possible any interleaving of actions from $P$ and $Q$, preserving all relations that exist in $P$ and in $Q$. Remember that when processes from other concurrency models are represented in the EAP model, any action consists of a single point-like event.

It is also possible to build the merge $P \parallel Q$ of processes $P$ and $Q$ by taking the free composition of all possible free interleavings of processes $P$ and $Q$.

The **right-merge** $P \rightmerge Q$ of processes $P$ and $Q$ is obtained from the free composition of processes $P$ and $Q$ in a similar way, only the compatibility relations make possible only interleavings of actions from $P$ and $Q$ that start with the first action from the process $Q$.

The **left-merge** $P \leftmerge Q$ of processes $P$ and $Q$ is obtained from the free composition of processes $P$ and $Q$ in a similar way, only the compatibility relations make possible only interleavings of actions from $P$ and $Q$ that start with the first action from the process $P$.

The **communication operator** is modeled (realized) in the EAP process algebra by means of special data transition actions. These actions take some results of actions in one process and give these results as inputs (initial conditions) to events and actions of another process.

To formalize communication in EAP, we need new constructions of input and output semantics, as well as a new operation of the strong sequential composition of events, actions and processes.

Let a structure $D$ be the common domain and a structure $C$ be the common codomain for a set events **E**.

If we interpret events as changes in some system $R$, then it is natural to interpret the common domain $D$ as the collection of possible states of the system $R$ before some event from **E** happens, and to interpret the common codomain *C* as the collection of possible states of the system *R* after some event from **E** occurs. For instance, we can take *D* equal to *C* and equal to the set St(*R*) of all states of *R*. When *R* is a distributed system that consists of subsystems $R_i$ ($i \in I$), then the set *D* = *C* can be equal to the union $\bigcup_{i \in I} St(R_i)$ or to the direct (Cartesian) product $\prod_{i \in I} St(R_i)$ where $St(R_i)$ is the set of all states of the system $R_i$.

**Definition 4.1.** The input semantics *InSem* is a partial mapping *InSem*: **E** → *D* and the output semantics *OutSem* is a partial mapping *OutSem*: **E** → *C*.

Input and output semantics form the base for functional programming languages, as well as for the logical approach to program verification (Floyd, 1959; Naur, 1966; Hoare, 1969).

**Definition 4.2.** The value *InSem*(*e*) is interpreted as input (the initial state) for an event *e* and the value *OutSem*(*e*) is interpreted as input (the initial state) for an event *e*.

The case when $D = C = \bigcup_{i \in I} St(R_i)$ gives a local input/output semantics for events where each event changes only one subsystem $R_i$.

The case when $D = C = \prod_{i \in I} St(R_i)$ gives a global input/output semantics for events where each event can change many subsystems $R_i$ and change in one of these subsystems is considered as change in the whole system *R*.

Input and output semantics are used in different disciplines and sometimes completely represent events. For instance, in probability theory, an event is a collection of possible outcomes of an experiment. Individual outcomes comprising an event are said to be *favorable* to that event.

A terminology more consistent with reality assumes that an event is a possible trial or experiment, while all possible outcomes of this trial or experiment form the output semantics of this event. A possible trial or experiment is a potential event. An event is said to occur as a result of an experiment if it contains the actual outcome of that experiment. An experiment that occurred is an actualized event.

Transitions in finite automata, abstract state machines, higher dimensional automata (Goubault, 1993) and labeled transition systems (Sassone, et al, 1996) are events that are represented by their input semantics (the initial state of the transition) and output semantics (the final state of the transition).

**Definition 4.3.** An event $e$ is the *strong sequential composition* of events $e_1$ and $e_2$ if $InSem(e) = InSem(e_1)$, $OutSem(e) = OutSem(e_2)$, $InSem(e_2) = OutSem(e_1)$ and $p_T(e_1) < p_T(e_2)$.

It is denoted by $e_1 \circ e_2$.

Informally, the strong sequential composition of events $e_1$ and $e_2$ means that the result of the event $e_1$ is taken as the input to the event $e_2$. In some cases, e.g., when $e_1$ and $e_2$ are computational operations performed by different devices, the strong sequential composition demand communication (data transmission) between these devices.

**Definition 4.4.** An action $A$ is the *strong sequential composition* of actions $A_1$ and $A_2$ if all events in $A$ are strong sequential compositions of events from $A_1$ and $A_2$.

It is denoted by $A_1 \circ A_2$.

The strong sequential composition of actions is, in some sense, the opposite to the free sequential composition of actions. However, there are many compositions between these extreme cases.

**Definition 4.5.** An action $A$ is the *mixed sequential composition* of actions $A_1$ and $A_2$ if it consists of two types of events: some events from $A$ belong to the free sequential composition of $A_1$ and $A_2$, while all other events in $A$ are strong sequential compositions of events from $A_1$ and $A_2$.

It is denoted by $A_1 \lozenge A_2$.

It is possible to introduce similar compositions for processes.

**Definition 4.6.** An process $P$ is the *strong sequential composition* of processes $P_1$ and $P_2$ if all actions in $P$ are strong sequential compositions of actions from $P_1$ and $P_2$.

It is denoted by $P_1 \circ P_2$.

The strong sequential composition of actions is, in some sense, the opposite to the free sequential composition of actions. However, there are many compositions between these extreme cases.

**Definition 4.7.** A process $P$ is the *mixed sequential composition* of processes $P_1$ and $P_2$ if it consists of two types of actions: some actions from $P$ belong to the free sequential composition of $P_1$ and $P_2$, while all other actions in $P$ are strong sequential compositions of actions from $P_1$ and $P_2$.

It is denoted by $P_1 \lozenge P_2$.

This definition shows that the traditional sequential composition of operators, automata or algorithms needs communication between the first and the second composed operators (automata or algorithms).

Now we can see that the communication operator from different process algebras is of the kind of mixed sequential composition $P_1 \lozenge P_2$ of processes in the EAP process algebra.

It is possible to represent the abstraction operator in the EAP process algebra as an operator of compression (shrinking) of the observability relation in processes. In some cases, compression of the observability relation is equivalent to deletion of some events in a process.

There is another way to model abstraction. Namely, it is possible to introduce silent step (unobservable) events and change some events in actions and processes for silent step (unobservable) events.

Note that constructive realization of the alternative operator demands an additional action that makes choice between $P$ and $Q$, may be under some condition.

As time and space are represented in EAP with much more detail that in other concurrency models, representation of operators from other models, such as merge or communication, is not unique in the EAP model.

The alternative operator and sequential composition define the *basic process algebra* ACP, which satisfies the following axioms (cf., for example, ((Bergstra and Klop, 1984)).

*Commutativity of the alternative operator*:  $x + y = y + x$

*Associativity of the alternative operator*:  $(x + y) + z = x + (y + z)$

*Idempotency of the alternative operator*:  $x + x = x$

*Distributivity of the sequential composition with respect to the alternative operator*:

$$(x + y) \cdot z = (x \cdot z) + (y \cdot z)$$

*Associativity of the sequential composition*: $(x \cdot y) \cdot z = x \cdot (y \cdot z)$

It is possible to show that the described above realizations in the EAP process algebra of the alternative operator and sequential composition preserves all these identities. Thus, we obtain a monomorphism (faithful representation) of the process algebra ACP in the EAP process algebra.

## 5. Conclusions and Future Work

The constructed algebra of concurrent processes we presented has a multilayer structure. These layers are divided into three levels. The algebra of events forms the first, lower level of the algebra of processes, while the algebra of actions forms the second, intermediate level of the algebra of processes. The third, highest level is formed by operations over whole processes. Each level can consist of several layers depending on the exactness of approximation and model precision.

When construction of the EAP process algebra is finished, the next step will be to develop a logical calculus for reasoning about processes and operations with them. The introduced constructions and obtained results imply several directions for future research.

First, it would be interesting to further develop concurrent programming based on the EAP model, to study concurrent programs, and to use such programs for developing software and firmware for computer networks and embedded networks (Manna and Pnueli, 1986).

It is important to note that concurrent programming is closely related to, and must utilize, superrecursive algorithms (Burgin, 2005). As it is emphasized in (Emerson, 1990), continuously operating concurrent programs exhibit ongoing behavior that is ideally non-terminating.

Next, an important problem is to build concurrent schemas based the EAP model, to study their properties, and to use such schemas for developing hardware, software and firmware for computer networks and embedded networks, as well as for finding ways to improve utilization of such networks. Mathematical tools for representation and study of such schemas are given in the mathematical schema theory (Burgin, 2005a; 2006; 2006a). Some types of which were introduced and studied in (Mazurkiewicz, 1977).

Finally, it is important to develop exact mathematical tools for comparison of different process algebras. Theory of categories provides some of such tools. However, to compare individual algebras, we need special structures from universal algebra.